\documentclass[12pt]{article}
\usepackage{amsmath,amssymb}
\usepackage{amsfonts}

\renewcommand{\labelenumi}{\roman{enumi}.}
\newcommand{\restrict}{\!\!\upharpoonright\!\!}
\newcommand{\range}{\mathrm{Ran}\,}
\newcommand{\domain}{\mathrm{Dom}\,}

\newcommand{\orbit}[1]{{\mathcal O}_{#1}}

\newtheorem{theorem}{Theorem}

\newtheorem{condition}[theorem]{Condition}

\newtheorem{principle}[theorem]{Rule}
\newtheorem{definition}[theorem]{Definition}

\newtheorem{lemma}[theorem]{Lemma}

\newtheorem{proposition}[theorem]{Proposition}
\newtheorem{remark}[theorem]{Remark}

\newenvironment{proof}[1][Proof]{\noindent\textbf{#1.} }{\ \rule{0.5em}{0.5em}}
\input{tcilatex}
\begin{document}

\title{Open Systems Viewed Through Their Conservative Extensions}
\author{Alexander Figotin \\
University of California at Irvine \and Stephen P. Shipman \\
Louisiana State University at Baton Rouge}
\maketitle

\begin{abstract}
A typical linear open system is often defined as a component of a larger
conservative one. For instance, a dielectric medium, defined by its
frequency dependent electric permittivity and magnetic permeability is a
part of a conservative system which includes the matter with all its atomic
complexity. A finite slab of a lattice array of coupled oscillators
modelling a solid is another example. Assuming that such an open system is
all one wants to observe, we ask how big a part of the original conservative
system (possibly very complex) is relevant to the observations, or, in other
words, how big a part of it is coupled to the open system? We study here the
structure of the system coupling and its coupled and decoupled components,
showing, in particular, that it is only the system's unique minimal extension
that is relevant to its dynamics, and this extension often is tiny part of
the original conservative system. We also give a scenario explaining why
certain degrees of freedom of a solid do not contribute to its specific
heat.
\end{abstract}

\section{Introduction: Open systems and conservative extensions}

Our interest in open systems is motivated, as it often happens, by a few
concrete problems which have something in common. One of the most important
concerns a time-dispersive dissipative (TDD) dielectric medium and the fundamental
problem of defining and studying the eigenmodes and, more generally, the
spectral theory. Similar problems arise when considering an "open resonator" (or
Helmholtz resonator), which is a regular resonator coupled to an exterior
system with an absolutely continuos spectrum.
A third problem originates in statistical
mechanics, when one considers a finite cube or slab as a small part of an
ideal solid, modeled by a lattice array of coupled oscillators, and wonders
how much information can be extracted about the entire solid from
observations made only from within the finite part. The common feature of
above dynamical systems is that they are open, in other words, the time
dynamics do not preserve the energy and/or the material but exchange them
with an exterior which often is not observable. It turns out that it is
possible to reach some interesting conclusions about properties of open
systems based on their minimal conservative extensions introduced recently
in \cite{FS}. This is the subject of this paper.

As an indication of some of what lies ahead, we mention a ``toy" example of a mechanical system, which we discuss in Section \ref{sectionFrozen}.
\label{toy}  The example shows how a high degree of symmetry in the system, which is related to high spectral multiplicity of the governing operator, results in many motions of the system being unaffected by a coupling to another system of  ``hidden" variables.  
Indeed, comparisons of the computation of the specific heat of a crystalline solid to experiment indicate that certain motions, or degrees of freedom, of the structure have to be left out---they are ``frozen" \cite[Section 3.1]{Gallavotti}, \cite[Section 6.4]{Huang}.
One of our main theorems (Theorem \ref{thmMultiplicity}) applies to this type of situation---it describes how the multiplicity of the modes of a system that are affected by a coupling to another system is bounded by the rank of the coupling.
Another example of a possible application of this work is one that inspired us to begin the study, although we do not pursue it at this point.  It concerns a periodic dielectric waveguide that admits ``nonrobust" modes at certain isolated wave number and frequency pairs.  These are true (nonleaky) modes that become leaky through radiation loss under perturbation of the frequency or wavenumber.  The system of modes in the waveguide and the exterior system, characterized by extended states in the surrounding air, become coupled, and this coupling produces interesting transmission anomalies \cite{SV}.

One can think of two intimately related and complementing ways to define an
open system: (i) \emph{intrinsic description by a non-conservative evolution
equation}; (ii) as \emph{a subsystem of a conservative system}.
Taking the intrinsic description as basic we define an \emph{open system} as
one governed by a causal time-homogeneous linear evolution equation 
\begin{equation}
m\partial _{t}v\left( t\right) =-\mathrm{i}Av\left( t\right)
-\int_{0}^{\infty }a\left( \tau \right) v\left( t-\tau \right) \,\mathrm{d}%
\tau +f\left( t\right) ,\quad v\left( t\right) \in H_{1},  \label{ops1}
\end{equation}%
in which $H_{1}$ is a separable Hilbert space, $m$ and $A$ are
self-adjoint operators in $H_{1}$ with $m>0$, and $f\left( t\right) $ is an external
force in $H_{1}$. We always assume that \emph{the system is at rest for
negative times} $t\leq 0$, in other words the following \emph{rest condition}
is satisfied%
\begin{equation}
v\left( t\right) =0,\ f\left( t\right) =0\text{ for }t\leq 0.  \label{ops1d}
\end{equation}
The integral term in (\ref{ops1}) involving the operator-valued response function
$a\left( \tau \right) $ is a subject to the \emph{dissipation (no-gain) condition}
\begin{equation}
\mathrm{Re}\int_{0}^{\infty }\int_{0}^{\infty }\overline{v\left( t\right) }%
a\left( \tau \right) v\left( t-\tau \right) \,\mathrm{d}t\,\mathrm{d}\tau
\,\geq \,0\quad \text{for all }v(t)\text{ with compact support}.
\label{ops2}
\end{equation}%
Evidently, the integral term in (\ref{ops1}) is responsible for the
non-conservative, or open, nature of the system. Its form explicitly
accounts for the system's causality and time-homogeneity. The friction
function $a(t)$ represents both delayed response and instantaneous friction;
thus we take it to be of the form 
\begin{equation}
a(t)\,=\,a_{\infty }\delta (t)+\alpha (t),  \label{ops1a}
\end{equation}%
where the coefficient of \emph{instantaneous friction} $a_{\infty }$ is a
bounded non-negative operator in $H_{1}$ and the \emph{delayed response
function} $\alpha (t)$ is strongly continuous and bounded as an
operator-valued function of $t$ with respect the norm in $\mathcal{B}(H_{1})$, the space of bounded operators in $H_{1}$.

The other important view on an \emph{open system is that it is a subsystem} of
a given conservative (conservative) system $(\mathcal{H},\mathcal{A})$, described
by conservative evolution equation 
\begin{equation}
\mathcal{M}\partial _{t}\mathcal{V}\left( t\right) =-i\mathcal{AV}\left(
t\right) +\mathcal{F}\left( t\right) ,\quad \mathcal{V},\mathcal{F}\in 
\mathcal{H},  \label{ops1b}
\end{equation}
where $\mathcal{H}$ is a separable Hilbert space, $\mathcal{A}$ is a
self-adjoint operator in it, and $\mathcal{F}\left( t\right) $ is an
external force, with a subsystem identified by a subspace $H_{1}\subset 
\mathcal{H}$. We refer to the subsystem's space $H_{1}$ as the \emph{observable
variables}.

An intimate relation between the two ways of looking at open systems can be
described as follows, \cite{FS}, \cite{FS1}:

\begin{enumerate}
\item[(i)] an open system defined by (\ref{ops1}) and satisfying (\ref{ops2}) can always be represented as a subsystem of a conservative extension in
the form (\ref{ops1b}), and, \emph{if minimal, such an extension is unique}
up to isomorphism;

\item[(ii)] the evolution of a subsystem of a conservative system (\ref%
{ops1b}) can be represented in the form (\ref{ops1}) with a friction
function $a(t)$ satisfying (\ref{ops2}).
\end{enumerate}

More precisely, taking the subsystem point of view on an open system we can
identify an open system (\ref{ops1}) with \emph{a subsystem of its minimal
conservative extension} $(\mathcal{H},\mathcal{A})$ in which a subspace
$H_{1}\subset \mathcal{H}$ acts as the space of observable variables. Then we
define the open system's exterior as the orthogonal complement $H_{2}=%
\mathcal{H}\ominus H_{1}$, referring to it as the \emph{hidden variables}.
Having the decomposition $\mathcal{H}=H_{1}\oplus H_{2}$ we can recast the
evolution equation (\ref{ops1b}) into the following system (see \cite[%
Section 2]{FS}). 
\begin{gather}
m_{1}\partial _{t}v_{1}\left( t\right) =-\mathrm{i}Av_{1}\left( t\right)
-\mathrm{i}\Gamma v_{2}\left( t\right) +f_{1}\left( t\right) ,\ m_{1}>0,\
A\text{ is self-adjoint,}  \label{ops2a} \\
\partial _{t}v_{2}\left( t\right) =-\mathrm{i}\Gamma ^{\dagger }v_{1}\left(
t\right) -\mathrm{i}\Omega _{2}v_{2}\left( t\right) ,\,\Omega _{2}\text{ is
self-adjoint,}  \notag
\end{gather}%
where $v_{1}\in H_{1},\ v_{2}\in H_{2}$, and $\Gamma :H_{2}\rightarrow H_{1}$
is the \emph{coupling operator}. By its very form, the system (\ref{ops2a}),
involving the coupling operator $\Gamma $ and its adjoint $\Gamma ^{\dagger
}:H_{1}\rightarrow H_{2}$, is explicitly conservative, and $\Gamma^\dag$ and
$\Gamma$ determine the 
channels of ``communication" from the observable to the
hidden and back from the hidden to the observable. Observe that if one
solves the second equation in (\ref{ops2a}) for $v_{2}\left( t\right) $ and
inserts it into the first equation, the resulting equation
will be of the form (\ref{ops1}) with friction function%
\begin{equation}
a(t)=\Gamma \mathrm{e}^{-\mathrm{i}\Omega _{2}t}\Gamma ^{\dagger },\quad
t\geq 0,  \label{ops2b}
\end{equation}%
which, as is easy to verify, always satisfies the dissipation condition (\ref%
{ops2}).

In a typical example of an open system embedded within
a given conservative system, this conservative system is not necessarily minimal as a
conservative extension of the open system.  The minimal conservative extension is often much simpler system than the original one.
For instance, a time dispersive and
dissipative (TDD) dielectric medium, as described by the Maxwell equations
with frequency dependent electric permittivity $\varepsilon $ and magnetic
permeability $\mu$, constitutes an open system. Note that such $\varepsilon$ and
$\mu$ arise through the interaction of the electromagnetic
fields with the molecular structure of the matter, which plays the part of
the hidden variables. But if, however, $\varepsilon $ and $\mu $ are all
that is known, clearly these functions would not allow one to \emph{%
reconstruct} the full molecular structure of the matter but rather only its
minimal conservative extension.

Another simple but instructive example is provided by a general \emph{scalar
(one-dimensional) open system} as described by (\ref{ops1}) with $H_{1}=
\mathbb{C}$ and friction function $a\left( t\right)$ satisfying (\ref%
{ops2}) and (\ref{ops1a}). Observe that the classical damped oscillator with 
$a(t)\,=\,a_{\infty }\delta (t)$ is a particular case of such general scalar
open system. The minimal conservative extension of a general\ scalar open
system is described by a triplet $\left\{ H_2,\Omega _{2},\Gamma \right\} $ such that (\ref{ops2b}) holds and its elements 
$H_{2}$, $\Omega _{2}$ and $\Gamma$ are constructed as
follows, \cite[Section 4.1, 5.1, 5.2, A.2]{FS}. First, using the Bochner
Theorem, we obtain the following representation of the friction function 
\begin{equation}
a\left( t\right) =\int_{-\infty }^{\infty }e^{-\mathrm{i}\omega t}\,\mathrm{d}N\left( \omega \right)  \label{agam2}
\end{equation}%
with a unique, non-decreasing, right-continuous bounded function $N\left(
\omega \right) $ defining a nonnegative measure $N\left( \mathrm{d}\omega
\right) $ on the real line $\mathbb{R}$. Then 
\begin{equation}
H_{2}=L^{2}\left( \mathbb{R},N\left( \mathrm{d}\omega \right) \right) ,
\label{agam3}
\end{equation}%
the operator $\Omega _{2}$ is the multiplication by $\omega $ on $%
L^{2}\left( \mathbb{R},N\left( \mathrm{d}\omega \right) \right) $, i.e.%
\begin{equation}
\left[ \Omega _{2}\psi \right] \left( \omega \right) =\omega \psi
\left( \omega \right) ,\ \omega \in \mathbb{R},\ \psi \in L^{2}\left( 
\mathbb{R},N\left( \mathrm{d}\omega \right) \right) ,  \label{agam4}
\end{equation}%
and the coupling operator $\Gamma$ and its adjoint are 
\begin{eqnarray}
\Gamma \left[ \psi \left( \cdot \right) \right] = \int_{-\infty
}^{\infty }\psi \left( \omega \right) \,N\left(\mathrm{d}\omega \right)
&:& L^{2}(\mathbb{R},dN)\rightarrow \mathbb{C},  \label{agam5} \\
\left[ \Gamma ^{\dagger }v\right] \left( \omega \right) = v, && \ v\in \mathbb{C%
},\ \omega \in \mathbb{R}.  \notag
\end{eqnarray}%
Consequently, the minimal conservative extension of the form (\ref{ops2a})
becomes here%
\begin{gather}
m\partial _{t}v=-\mathrm{i}Av-\mathrm{i}\int_{-\infty }^{\infty }\psi \left(
\omega \right) \,d\omega +f\left( t\right) ,  \label{agam6} \\
\partial _{t}\psi \left( \omega \right) = -\mathrm{i}v-\mathrm{i}\omega \psi
\left( \omega \right) ,\ \psi \in L^{2}\left( \mathbb{R},N\left( \mathrm{d}%
\omega \right) \right) .  \notag
\end{gather}%
In the case of the classical damped oscillator the measure $N\left( \mathrm{d%
}\omega \right) $ is just the Lebesque measure, i.e. $N\left( \mathrm{d}%
\omega \right) =\mathrm{d}\omega $, and the system (\ref{agam6}) is
equivalent to the Lamb model (see \cite{Lamb} and \cite{FS1}), which is a point mass attached to a classical elastic string (with $\omega $ being the wave number). In the
case of a general spectral measure $N\left( \mathrm{d}\omega \right) $ one
can view the minimal extension (\ref{agam6}) as one obtained by attaching
a point mass $m$ to a general ``\emph{string}" as described by a simple,
i.e. multiplicity-one, self-adjoint operator with the spectral measure $%
N\left( \mathrm{d}\omega \right) $. This point of view is justified by\emph{a fundamental construction due to M. G. Krein of a unique ``real" string
corresponding to any given spectral measure}. This construction as a part of
an exhaustive study of relations between the spectral measure, the
corresponding admittance operator (the coefficient of dynamical compliance),
and strings, is presented in two papers \cite{KKr1,KKr2} by I. S. Kac
and M. G. Krein.

To clarify the exact meaning of a string, we give a
brief description of a loaded string $S_{1}\left[ 0,L\right] $ on an
interval $\left[ 0,L\right] $, $0\leq L\leq \infty $, as it is presented by Kac and Krein
\cite{KKr1,KKr2}.
We assume (i) the string $S_{1}\left[ 0,L\right] $
has constant stiffness $1$; (ii) a nondecreasing nonnegative function $%
M\left( s\right) $, $s\geq 0$, describes its mass distribution, with $%
M\left( s\right) $ being the total string mass on the interval $\left[ 0,s%
\right] $. The string states are complex-valued functions $\psi \left(
s\right) $, $0\leq s\leq L$, from the Hilbert space $L^{2}\left( \left[ 0,L%
\right] ,M\left( \mathrm{d}s\right) \right) $. The string dynamics is
governed by the following equation 
\begin{equation}
\frac{\partial ^{2}\psi }{\partial t^{2}}\left( s,t\right) =A_{M}\left[ \psi %
\right] \left( s,t\right) ,\ 0\leq s\leq L,  \label{kkr1}
\end{equation}%
where the string operator $A_{M}$ is defined by the expression 
\begin{equation}
A_{M}\left[ \psi \right] \left( s\right) =-\frac{d}{dM\left( s\right) }\frac{%
d\psi }{ds}\left( s\right) ,\ 0\leq s\leq L,  \label{kkr2}
\end{equation}%
with the boundary conditions%
\begin{equation}
\psi ^{\prime }\left( 0\right) =0,\ \psi ^{\prime }\left( L\right) h+\psi
\left( L\right) =0,\ \text{where }h\text{ is real.}  \label{kkr3}
\end{equation}%
We do not formulate the original statements from \cite[Theorem 11.1, 11.2]%
{KKr2}, because of the considerable space needed to introduce and define all
relevant concepts, but their principal point is that any nonnegative measure 
$N\left( \mathrm{d}\omega \right) $ on $\left( -0,\infty \right) $
satisfying the condition%
\begin{equation}
\int_{-0}^{\infty }\frac{N\left( \mathrm{d}\omega \right) }{1+\omega }<\infty
\label{kkr4}
\end{equation}%
is the spectral measure of a unique string as described by the self-adjoint
operator $A_{M}$ defined by (\ref{kkr2})-(\ref{kkr3}). In another words,
given a nonnegative measure $N\left( \mathrm{d}\omega \right) $ on the
positive semiaxis $\left( -0,\infty \right) $ which satisfies the condition (\ref{kkr4}), one can construct a unique mass distribution $M\left( s\right) $, so that the corresponding string $S_{1}\left[ 0,L\right] $ has $N\left(\mathrm{d}\omega \right) $ as its spectral measure. As to the relation
between $N\left( \mathrm{d}\omega \right) $ and $M\left( s\right) $ a number
of insightful examples are provided in \cite[Sections 11-13]{KKr2}.

Observe now that, if a scalar open system is described by (\ref{ops2a}), then 
\emph{regardless of how complex the original triplet
$\left\{H_{2},\Omega _{2},\Gamma \right\}$ is, its minimal counterpart
$\left\{H_{2,\min },\Omega _{2,\min },\Gamma _{\min }\right\}$ is always of
the universal form (\ref{agam6}), and one can think of it as obtained by
attaching a string to a point mass.} The concept of a string, as represented
by the spectral measure $N\left( \mathrm{d}\omega \right) $ on $\mathbb{R}$,
turns out to be useful in describing the minimal extension of a
multidimensional open system, where one has to use a number of strings for
its construction. In particular, we will introduce a rather simple \emph{%
string spectral decomposition} for an arbitrary self-adjoint operator
$\Omega_2$ for which the number of strings equals exactly to the spectral
multiplicity of $\Omega_2$, and \emph{a single string has always spectral
multiplicity one}. We use then the number of strings involved in the string
decompositions to characterize their relative complexity.

We reiterate the important observation that follows from the above examples and discussion: \emph{the evolution of a subsystem is fully described by its minimal conservative extension similar to the system (\ref{agam6}) which, typically, is
substantially simpler than the original conservative system}. Consequently,
a significant part of the modes of the original system can be completely
decoupled from the open system. These observations make the minimal
conservative extension an attractive instrument: (i) it is a simpler
substitute for often enormously complex original conservative systems (as
the atomic structure of the matter), (ii) since it is conservative, the
classical spectral theory is available, and (iii) it provides information
about how much of the original system is reconstructible by an observer in
the open subsystem.  Based on our considerations hitherto, we see our objectives
as follows:

\begin{enumerate}
\item[(i)] identify information about a conservative system that is carried
by its subsystem (\emph{reconstructibility});

\item[(ii)] relate the unique minimal extension of an open system to a given
``original" larger conservative system;

\item[(iii)] study the coupling operator of a subsystem and to identify
which part of the system is coupled through it.

\item[(iv)] understand the decomposition of open systems through
simultaneous decompositions of the internal dynamics of the observable and
hidden variables and the coupling operator between them.
\end{enumerate}

{\bfseries Organization of the paper.} \ 
Section \ref{secExtension} gives precise definitions and concise mathematical discussions of the concepts introduced so far, which will serve as background for the development of the work.

In Section \ref{sectionFrozen}, we construct a toy model of a solid that has frozen degrees of freedom.  This example serves to illustrate the role that a high degree of system symmetry plays in decoupling parts of a dynamical system, as we have already discussed (page \pageref{toy}).

Section \ref{sectionReconstructibility} concerns the \emph{reconstructibility} of conservative systems from open systems, in particular, from the dynamics projected to the observable and to the hidden state variables.  The section culminates in one of our main theorems, Theorem \ref{thmMultiplicity}, which describes how the number of coupling channels between the observable and hidden variables bounds the number of strings required in the construction of the minimal extension.

In Section \ref{sectionDecomposition}, we
investigate the decomposition, or decoupling, of open systems by means of
the minimal conservative extension. We show first the equivalence between
(i) the decoupling of the dynamics in a subspace of the open system from the
dynamics in the complementary part of the open system, which is determined
by $A_{1}$ and $a(t)$, and (ii) splittings of the conservative extension
that are invariant under $A_{1}$, $\Omega _{2}$, and $\Gamma $, or,
equivalently, that are preserved in $\mathcal{H}$ by $\Omega $ and the
projection operator to $H_{1}$.
We then make an analysis of the relation between splittings of the
conservative extension $\mathcal{H}$ that preserve its dynamics
(equivalently, splittings of the projected open system on $H_{1}$) and the
singular-value decomposition of the coupling operator $\Gamma $. If the
friction function involves no instantaneous friction component, that is, if $%
a_{\infty }=0$, then the coupling $\Gamma $ is bounded; otherwise it will be
unbounded. More general assumption when $\Gamma $ is bounded with respect to
the frequency operator $\Omega _{2}$, which covers the case of nonzero
instanteneous fiction $a_{\infty }$, is considered in \cite[Section 2.2]{FS}%
. In this work we focus on the case of bounded coupling, in which the
analysis is more transparent.

To preserve the conceptual transparency of the arguments and results and to
make them more readily accessible to the reader, we forgo full
rigorous arguments in the development of the ideas and present more
elaborate statements of the theorems as well as their proofs in Section
\ref{sectionProofs}.

\section{Conservative extension and spectral composition}

\label{secExtension}

Our study of the general open linear DD system is based on the ability to
embed it in a unique way into a larger conservative system, in which the
observable system is complemented by a space of hidden degrees of
freedom. The frequency operator for the hidden variables gives rise to a
(nonunique) decomposition of these variables into subspaces that are
interpreted as independent ``strings" that are ``attached" to the system of
observable variables and account for the dissipative and dispersive effects
that cause this system to be open. Each string is characterized by a
spectral measure, and exactly how they strings are attached to the
observable variables is described by the coupling operator. In this section,
we give the background for constructing this conservative extension and
discuss the spectral composition of the space of hidden variables into strings and the structure of the coupling operator.

\medskip

We reiterate our definition of an open linear DD system and the conditions it satisfies.
We take an open system to be of the form
\begin{equation}
m\partial _{t}v\left( t\right) =-\mathrm{i}Av\left( t\right)
-\int_{0}^{\infty }a\left( \tau \right) v\left( t-\tau \right) \,\mathrm{d}%
\tau +f\left( t\right) ,\quad v\left( t\right) \in H_{1},  \label{opensystem}
\end{equation}
in which $H_{1}$ is a separable Hilbert space, $m$ and $A$ are
self-adjoint operators in $H_{1}$ with $m>0$, and $f\left(t\right)$ is an external
force in $H_{1}$.
The function $a(t)$ is subject the dissipation (no-gain) condition

\begin{condition}[dissipation] \label{condDissipation}
Let $a(t)=a_{\infty }\delta (t)+\alpha (t)$, where $a_{\infty }$ is a
bounded non-negative operator in $H_{1}$ and $\alpha(t)$ is a strongly continuous and bounded operator-valued function of $t$ with respect the operator norm in
$\mathcal{B}(H_{1})$.  $a(t)$ satisfies the \emph{dissipation condition} if
\begin{equation}
\mathrm{Re}\int_{0}^{\infty }\int_{0}^{\infty }\overline{v\left( t\right) }%
a\left( \tau \right) v\left( t-\tau \right) \,\mathrm{d}t\,\mathrm{d}\tau
\,\geq \,0\quad \text{for all }v(t)\text{ with compact support}.
\label{dissipation}
\end{equation}
\end{condition}

\noindent
The systems we consider will satisfy the rest condition

\begin{condition}[rest condition] \label{condRest}
An open system satisfies the \emph{rest condition} if, for all $t<0$, $f(t)=0$ and
$v(t)=0$.
\end{condition}

{\bfseries The minimal extension.} \ The following statement, which is a
generalization of the Bochner theorem, plays the key role in the embedding
of the open system \eqref{ops1} into a unique minimal conservative extension %
\eqref{ops2a} \cite[Theorem 3.2]{FS}. Given that the dissipation
condition \eqref{ops2} is satisfied, the Proposition \ref{propExtension}
provides the existence of the space $H_2$ of hidden variables, the frequency
operator $\Omega_2$ for its internal dynamics, and the coupling
operator~$\Gamma$.

\begin{proposition}[mimimal extension]
\label{propExtension} \label{pmin1}Let $\mathcal{B}\left( H_{1}\right) $ be
the space of all bounded linear operators in $H_{1}$. Then a strongly
continuous $\mathcal{B}\left( H_{1}\right) $-valued function $a\left(
t\right) $, $0\leq t<\infty $, is representable as%
\begin{equation}
a\left( t\right) =\Gamma e^{-\mathrm{i}t\Omega _{2}}\Gamma ^{\dagger },
\label{hg1b}
\end{equation}%
with $\Omega _{2}$ a self-adjoint operator in a Hilbert space $H_{2}$ and $%
\Gamma :$ $H_{1}\rightarrow H_{2}$ a bounded linear map, if and only if $%
a\left( t\right) $ satisfies the dissipation condition (\ref{ops2}) for
every continuous $H_{1}$ valued function $v(t)$ with compact support. If the
space $H_{2}$ is minimal---in the sense that the linear span%
\begin{equation}
\left\langle f\left( \Omega _{2}\right) \Gamma ^{\dagger }v:f\in C_{c}\left( 
\mathbb{R}\right) ,\text{ }v\in H_{1}\right\rangle  \label{hg1c}
\end{equation}%
is dense in $H_{2}$---then the triplet $\left\{ H_{2},\Omega _{2},\Gamma
\right\} $ is determined uniquely up to an isomorphism.
\end{proposition}

\begin{remark}
In fact, it is sufficient to assume that $a(t)$ is locally
bounded and strongly measurable, strong continuity then follows from (\ref%
{hg1b}).
\end{remark}

Since, as it turns out, spans similar to (\ref{hg1c}) arise often in the
analysis of open systems, we name them closed \emph{orbits} and define them as
follows.

\begin{definition}[orbit]
\label{dorbit}Let $\Omega$ be a self-adjoint operator in a Hilbert space $H$
and $S$ is a subset of vectors in $H$. Then we define the \emph{closed orbit} (or simply \emph{orbit}) \emph{${\mathcal{O}}_{\Omega }(S)$ of $S$ under action of $\Omega$} by 
\begin{equation}
{\mathcal{O}}_{\Omega }(S)=\text{closure of span}\left\{ f(\Omega )w:f\in
C_{c}(\mathbb{R}),\ w\in S\right\} .
\end{equation}
If $H^{\prime }$ is a subspace of $H$ such that ${\mathcal{O}}%
_{\Omega}(H^{\prime }) = H^{\prime }$, then $H^{\prime }$ is said to be 
\emph{invariant with respect to $\Omega$} or simply $\Omega$-invariant.
\end{definition}

If $\Omega $ is bounded, the orbit ${\mathcal{O}}_{\Omega }(S)$ is equal to
the smallest subspace of ${H}$ containing $S$ that is invariant, or closed,
under $\Omega $. Equivalently, it is the smallest subspace of ${H}$
containing $S$ that is invariant under $(\Omega -i)^{-1}$; this latter
formulation is also valid for unbounded operators. The relevant theory can
be found, for example, in \cite{AkhGlaz} or \cite{RS1}. The orbit of $S$
under the of two self-adjoint operators $\Omega$ and $A$ can be defined by
application of continuous functions of $\Omega$ and $A$ to elements of $S$,
but we shall only need the characterization that 
\begin{multline} \label{dorbit2}
{\mathcal{O}}_{\Omega,A}(S) \text{ is the smallest subspace of $H$
containing $S$} \\
\text{that is invariant under $(\Omega -i)^{-1}$ and $(A-i)^{-1}$.}
\end{multline}

Proposition \ref{pmin1} allows one uniquely to construct the triple $\left\{
H_{2},\Omega _{2},\Gamma \right\} $ and, consequently, the minimal
conservative extension based on the observable friction function $a(t)$. In fact, there is a statement similar to Proposition
\ref{pmin1} which holds for $a\left( t\right) $ of the most general form (%
\ref{ops1a}), in which instantaneous friction is included; see \cite[Theorem
7.1]{FS}. Consideration of the instantaneous friction term leads to an
unbounded coupling operator $\Gamma$; the treatment of unbounded coupling is
technical, and we do not consider it in this work, but treat it in a
forthcoming exposition.

\smallskip

It is worthwhile to understand the idea behind the construction of the
triple $\left\{ H_{2},\Omega _{2},\Gamma \right\}$, as it shows plainly how
the time-harmonic decomposition of $a(t)$ determines the spectral structure
of $H_2$. We therefore take a page to explain it. Introduce the
Fourier-Laplace transform of $a(t)$: 
\begin{equation}
\hat a(\zeta) = \int_0^\infty a(t) e^{i\zeta t}\, dt, \quad \text{for $%
\Im\zeta>0$}.
\end{equation}
It turns out that the dissipation condition, Condition \ref{condDissipation}, on $a(t)$ is equivalent
to the condition that $\hat a(\zeta)$ is a Nevanlinna function: it is an
analytic function of the open upper half plane with values that have
positive real (self-adjoint) part. The restriction of the real part of $\hat
a(\zeta)$ to the real line ($\hat a(\omega)$ for $\omega\in\mathbb{R}$ is
the Fourier transform of $a(t)$) is no longer a classical function in
general, but rather a \emph{nonnegative} operator-valued measure $dN(\omega)$%
. One then takes $H_2$ to be the space of square-integrable functions from $%
\mathbb{R}$ to $H_1$ with respect to this measure: 
\begin{equation}
H_2 = L^2(\mathbb{R}, H_1, dN(\omega)),
\end{equation}
for which the inner product is defined by 
\begin{equation}
\langle f | g \rangle_{H_2} = \frac{1}{\pi}\int_{\mathbb{R}} \langle
f(\omega) | dN(\omega) \, g(\omega) \rangle_{H_1},
\end{equation}
with the integral understood in the Lebesgue-Stieltjes sense. The operator $%
\Omega_2$ is simply multiplication by $\omega$: 
\begin{equation}
(\Omega_2(g))(\omega) = \omega g(\omega) \quad \text{for $g\in H_2$},
\end{equation}
the adjoint $\Gamma^\dag:H_1\to H_2$ of the coupling operator is defined by sending $v\in H_1$ to the function with constant value $v$: 
\begin{equation}
(\Gamma^\dag (v))(\omega) = v \quad \text{for all $\omega\in\mathbb{R}$}
\end{equation}
and $\Gamma:H_2\to H_1$ is given by 
\begin{equation}
\Gamma (f) = \frac{1}{\pi}\int_{\mathbb{R}} f(\omega) \, dN(\omega).
\end{equation}
One can check that this extension indeed produces the friction function $%
a(t) $ by observing that, since $\hat a(\zeta)$ is a Nevanlinna function, it
is constructible from $dN(\omega)$ by the Cauchy transform: 
\begin{equation}
i\hat a(\zeta) = \frac{1}{\pi} \int_\mathbb{R} \frac{1}{\zeta-\omega}
dN(\omega),
\end{equation}
and when applied to a vector $v\in H_1$, gives 
\begin{equation}
i\hat a(\zeta) v = \frac{1}{\pi} \int_\mathbb{R} \frac{1}{\zeta-\Omega_2}%
(\Gamma^\dag(v)(\omega)) dN(\omega) = \left( \Gamma \frac{1}{\zeta-\Omega_2}
\Gamma^\dag \right) v,
\end{equation}
which is the Fourier-Laplace transform of $(\Gamma e^{-i\Omega_2
t}\Gamma^\dag)v$.

\medskip

{\bfseries The coupling channels.} \ Now let us understand the structure of
the coupling operator $\Gamma$ well. There is a canonical isomorphism
between the ranges of $\Gamma$ and $\Gamma^\dag$: 
\begin{equation}  \label{U}
U:\mathrm{Ran}\,\Gamma \rightarrow \mathrm{Ran}\,\Gamma^\dagger.
\end{equation}
This isomorphism is constructed as follows: Observe that 
\begin{equation}
H_{1}=\overline{\mathrm{Ran}\,\Gamma }\oplus \mathrm{Null}\,\Gamma ^{\dagger
}\quad \text{and}\quad H_{2}=\overline{\mathrm{Ran}\,\Gamma ^{\dagger }}%
\oplus \mathrm{Null}\,\Gamma
\end{equation}
so that $\Gamma $ and $\Gamma ^{\dagger }$ are determined by their actions
on $\mathrm{Ran}\,\Gamma ^{\dagger }$ and $\mathrm{Ran}\,\Gamma $,
respectively. Denote their restrictions to these subspaces (both in domain
and target space) by 
\begin{equation}
\Gamma _{R}=\Gamma \!\!\upharpoonright \!\!\mathrm{Ran}\,(\Gamma ^{\dagger
})\quad \text{and}\quad \Gamma _{R}^{\dagger }=\Gamma ^{\dagger
}\!\!\upharpoonright \!\!\mathrm{Ran}\,(\Gamma )=(\Gamma _{R})^{\dagger }.
\end{equation}
$U$ is then given explicitly by 
\begin{equation}\label{UU}
U=\left( \Gamma _{R}^{\dagger }\Gamma _{R}\right) ^{-1/2}\Gamma
_{R}^{\dagger }=\Gamma _{R}^{\dagger }\left( \Gamma _{R}\Gamma _{R}^{\dagger
}\right) ^{-1/2}.
\end{equation}
%
%
The positive operators $(\Gamma _{R}\Gamma _{R}^{\dagger })^{1/2}$ on $%
\mathrm{Ran}\,\Gamma $ and\ $(\Gamma _{R}^{\dagger }\Gamma _{R})^{1/2}$ on $%
\mathrm{Ran}\,\Gamma ^{\dagger }$ have trivial nullspace and are related
through $U$ by\footnote{%
It is not always necessary to deal with the restrictions $\Gamma _{R}$ or $%
\Gamma _{R}^{\dagger }$; often $\Gamma $ or $\Gamma ^{\dagger }$ itself is
suitable. For example, $\Gamma \Gamma ^{\dagger }$ and $\Gamma
_{R}\Gamma_{R}^{\dagger }$ coincide on the domain of the latter, and the
former maps the orthogonal complement of this domain to zero. The analogous
statement holds for $\Gamma ^{\dag }\Gamma $ and $\Gamma _{R}^{\dag }\Gamma
_{R}$. \, $U^{-1}:\mathrm{Ran}\,{\Gamma^\dag}\to\mathrm{Ran}\,\Gamma$ is
given by $U^{-1}=U^{\dagger }= \left( \Gamma _{R}\Gamma _{R}^{\dagger
}\right)^{-1/2}\Gamma _{R}=\Gamma _{R} \left( \Gamma _{R}^{\dagger
}\Gamma_{R}\right) ^{-1/2}$, and $\Gamma$ has the polar decompositions $%
\Gamma_R = U^{-1}\left( \Gamma_{R}^{\dagger }\Gamma _{R}\right)^{1/2} =
\left( \Gamma _{R}\Gamma _{R}^{\dagger }\right)^{1/2}U^{-1}$.} 
\begin{equation}  \label{PolarDecomposition}
\Gamma_R^\dag = U \left( \Gamma _{R}\Gamma _{R}^{\dagger }\right)^{1/2} =
\left( \Gamma_{R}^{\dagger }\Gamma _{R}\right)^{1/2} U.
\end{equation}
This is the polar decomposition of $\Gamma^\dag$. It allows one to define
the ``coupling channels" in a natural way as the pairing of the eigenmodes
of the positive part $( \Gamma _{R}\Gamma _{R}^{\dagger })^{1/2}$ in $%
\mathrm{Ran}\,\Gamma\subset H_1$ with the corresponding eigenmodes of $(
\Gamma_{R}^{\dagger }\Gamma _{R})^{1/2}$ in $\mathrm{Ran}\,\Gamma^\dag%
\subset H_2$ through $U$. $\Gamma^\dag$ and $\Gamma$ provide a direct
coupling between these modes---hence the term ``coupling channel". In the
case of unbounded coupling or continuous spectrum, the modes are not genuine
vectors, but are members of a appropriate furnishings of $H_1$ and $H_2$.
For continuous spectrum, we may also define coupling channels more generally
as pairs of spaces identified through $U$ that are fixed by the positive
operators in \eqref{PolarDecomposition} and therefore mapped to one another
by $\Gamma^\dag$ and $\Gamma$. We use this structure amply in Section \ref%
{sectionDecomposition}, which deals with decomposition of open systems.

\begin{definition}[coupling channel]
\label{defCouplingChannel} A \emph{coupling channel} is a pair $(S_1,S_2)$,
in which $S_1$ is an invariant space of $( \Gamma _{R}\Gamma _{R}^{\dagger
})^{1/2}$ in $\mathrm{Ran}\,\Gamma\subset H_1$ and $S_2$ is an invariant
space of $( \Gamma_{R}^{\dagger }\Gamma _{R})^{1/2}$ in $\mathrm{Ran}%
\,\Gamma^\dag\subset H_2$ such that $U(S_1) = S_2$ (see \eqref{U} and \eqref{UU} for the definition of $U$). If follows that $%
\Gamma^\dag(S_1) = S_2$ and $\Gamma(S_2) = S_1$. A \emph{simple coupling
channel} is a coupling channel in which the members $S_1$ and $S_2$ are one-dimensional. A simple coupling channel corresponds to a pair of eigenmodes
$(\phi_1,\phi_2)$ of $\Gamma\Gamma^\dag$ and $\Gamma^\dag\Gamma$ for the same eigenvalue.
\end{definition}

\medskip

{\bfseries The extending strings.} \ The conservative system $(H_2,\Omega_2)$,
consisting of the space of hidden variables together with its operator of
internal dynamics, can be interpreted as a set of independent abstract
``strings" to which the system $(H_1,m,A)$ is attached by the coupling
channels defined by $\Gamma$.  The following spectral decomposition of $H_2$ with respect to $\Omega_2$ is obtained by a straightforward modification of Theorem VII.6 in \cite{RS1}:
\begin{equation}  \label{StringDecomposition}
H_2 \cong \bigoplus_{j=1}^M L^2(\mathbb{R},\mathbb{C},d\mu_j(\omega)), \quad
d\mu_{j+1} \preceq d\mu_j,
\end{equation}
in which ``$\preceq$" denotes absolute continuity of measures and $\Omega$ is
represented by multiplication by the independent variable $\omega$. We call
each component of this decomposition a ``string"; the $j$-th string is
generated by a function $f_j(\omega)$ of maximal spectral type in $L^2(%
\mathbb{R},\mathbb{C},d\mu_j(\omega))$, that is, $f_j(\omega)\not=0$ almost
everywhere with repect to $d\mu_j$. A string is characterized by its
invariance under the action of $\Omega_2$ and by the property that the
restriction of $\Omega$ to the string has multiplicity $1$. Of course, the
measures $\mu_j$ need not be taken to be nested by absolute continuity; even
if they are, a decomposition into strings is not unique. The strings are
decoupled from each other with respect to the action of $\Omega_2$, that is,
within the conservative system $(H_2,\Omega_2)$.

\begin{definition}[string]
\label{defString} An \emph{abstract string}, or simply a \emph{string} in
the system $(H_2,\Omega_2)$, is a subsystem $(S,\Omega_2\restrict S)$, in which
$S$ is a $\Omega_2$-invariant subspace of $H_2$ and the restriction $\Omega_2\!\!\upharpoonright\!\! S$ of $\Omega_2$ to $S$ has multiplicity~$1$.    A {\em string decomposition} of $(H_2,\Omega_2)$ is an expression of $(H_2,\Omega_2)$ as a direct sum of strings:
\begin{equation}\label{StringDecomposition2}
H_2 = \bigoplus_{j=1}^M H_{2j}\,, , \quad
\Omega_2 = \bigoplus_{j=1}^M \Omega_2\restrict{H_{2j}} \,,
\end{equation}
in which each $(H_{2j},\Omega_2\restrict{H_{2j}})$ is a string in $(H_2,\Omega)$.
\end{definition}

Evidently, the isomorphism \eqref{StringDecomposition} gives a string decomposition of $H_2$.  In view of the corresponding representation of $\Omega_2$ as multiplication by the independent variable $\omega$, construction of a decomposition of $H_2$ into strings is accomplished abstractly as follows: Choose a vector $v_1\in H_2$ of maximal spectral type with respect to $\Omega_2$ and obtain $H_{21} ={\mathcal{O}}_{\Omega_2}(v_1) \cong L^2(\mathbb{R},\mathbb{C},d\mu_1(\omega))$. Then, if ${\mathcal{O}}_{\Omega_2}(v_1)\not=H_2$, choose a vector $v_2$ of maximal spectral type in ${\mathcal{O}}_{\Omega_2}(v_1)^\perp$ and obtain $H_{22} = {\mathcal{O}}_{\Omega_2}(v_2) \cong L^2(\mathbb{R},\mathbb{C},d\mu_2(\omega))$, and so on.  This infinite iterative process will produce a direct sum of the form
\eqref{StringDecomposition2}.  However, if the vectors $v_n$ are chosen at will, this sum may not be all of $H_2$: it may have an orthogonal complement, within $H_2$, in which $\Omega_2$ has uniform infinite multiplicity.  One must be sure to include this part in the string decomposition.  The structure provided by
\eqref{StringDecomposition} shows that this is indeed possible.

The number $M$ (which may be infinite) in the spectral representation \eqref{StringDecomposition} is the multiplicity of the operator $\Omega_2$; $M$ is the maximal multiplicity of any of the spectral values of $\Omega_2$.

\medskip

{\bfseries Discussion.} \
The coupling of the observable variables $H_1$ to the strings is
accomplished through the coupling channels defined by $\Gamma$. Of course,
there is in general no relation between a given decomposition of $H_2$ into strings and the coupling channels. If the strings can be chosen in such a way that the coupling channels split into two sets, one of which couples into one set of strings and the other of which couples into the complementary set of strings, {\em and} the $H_1$-members of the two sets of channels are contained in orthogonal orbits of $\Omega_1$, then the {\em open} system $(H_1,\Omega_1,a(t))$ is decomposed into decoupled systems. We pursue a detailed study of the decoupling of open systems using their conservative extensions in Section~\ref{sectionDecomposition}.

In a typical example in which it is known that the open system $(H_1,m,A_1,a(t))$ is obtained naturally as the restriction of the dynamics of a given larger conservative system $(\mathcal H,\Omega)$ to a subspace of observable variables $H_1\subset \mathcal H$ (as the open system of electromagnetic fields in a lossy medium or a crystalline solid in contact with a heat bath), the given conservative system is not necessarily minimal.  The space of hidden variables for the minimal extension is actually a subspace of the given $H_2 = \mathcal H \ominus H_1$.  We call this subspace the ``coupled" part of $H_2$ and denote it by $H_{2c}$.  $H_{2c}$ coincides with $H_2$ if $(\mathcal H,\Omega)$ is minimal.

One of our main results concerns the situation in which there exist only finitely many simple coupling channels.  This is the case that the rank of $\Gamma$ is finite, such as in lattice systems, as we discuss in some detail as a motivating example in the following section.  The result  gives quantitative information about the size of $H_{2c}$ within $H_2$.  As $H_1$ acts as the ``hidden" variables for a hypothetical observer in $H_2$, we have also an analogous result about the size of $H_{1c}$ within $H_1$, where $H_{1c}$ is the part of $H_1$ that is reconstructible from the dynamics restricted to $H_2$ (or $H_{2c}$):

\vspace*{2ex}

\noindent
\hspace*{0.05\textwidth}
\parbox{0.9\textwidth}{\em
The (minimal) number of strings needed to extend an open system to a conservative one is no greater than the number of independent simple coupling channels between the spaces of observable and hidden variables.

\vspace*{1ex}
\noindent The coupled part of the observable variable space has multiplicity (with respect to $\Omega_1$) that is no greater than the number of independent simple coupling channels between the spaces of observable and hidden variables.}

\vspace*{2ex}

\noindent
This result is stated precisely as Theorem \ref{thmMultiplicity} in Section~\ref{sectionReconstructibility}, and its sigificance is discussed in Section~\ref{sectionFrozen}.

\section{Open systems and frozen degrees of freedom}
\label{sectionFrozen}

We illustrate through a quite concrete example that certain degrees of freedom of a DD system can be ``frozen":  they are not affected by the interaction with the hidden variables that causes the energy-dissipation effects.  Thus a component of the state space that is ``decoupled" from the hidden variables evolves conservatively, independent of the ``coupled" DD part.  This section may serve as a motivation for our detailed study in Section \ref{sectionReconstructibility}.

Consider a crystalline solid in contact with a heat bath. It has been
observed that certain degrees of freedom of the solid do not contribute to
its specific heat \cite[Section 3.1]{Gallavotti}, \cite[Section 6.4]{Huang}.
The calculation of the specific heat by the Dulong-Petit law is
based on the law of equipartition of energy and the number of degrees of
freedom.  For that calculation to agree with the experiment, one has to leave
out some degrees of freedom as if they were ``frozen" and
cannot be excited by the heat bath. In other words, there are system
motions which are completely decoupled from the solid and heat bath
interaction---they cannot be reached through the combination of surface
contact and internal dynamics of the solid.

To find a sufficiently general scenario for such frozen degress of freedom we
consider an open system decribed by the variable $v_{1}\in H_{1}$ as a part
of the conservative system (\ref{ops2a}). We notice then that it is
conceivable that the open system has a part not coupled to its exterior. In
other words, there is an orthogonal decomposition%
\begin{equation}
H_{1}=H_{1c}\oplus H_{1d},  \label{hhd1}
\end{equation}%
where the subspaces $H_{1c}$ and $H_{1d}$ correspond to states
coupled to and decoupled from the hidden variable $v_{2}\in H_2$. To figure out the
decompositon (\ref{hhd1}) we set $m_{1}=1$\ in (\ref{ops2a}) (the general
case is reduced to this one by proper renormalization of $v_{1}$), and
consider the system 
\begin{gather}
\partial _{t}v_{1}\left( t\right) =-\mathrm{i}\Omega _{1}v_{1}\left(
t\right) -\mathrm{i}\Gamma v_{2}\left( t\right) ,\ \Omega _{1}=\Omega
_{1}^{\dag }  \label{hhd2} \\
\partial _{t}v_{2}\left( t\right) =-\mathrm{i}\Gamma ^{\dagger }v_{1}\left(
t\right) -\mathrm{i}\Omega _{2}v_{2}\left( t\right) +f_{2}\left( t\right) ,\
\Omega _{2}=\Omega _{2}^{\dag }.  \notag
\end{gather}%
The system (\ref{hhd2}) allows one to single out states $v_{1}$ which can be
excited by the variables $v_{2}$, which constitute subspace $H_{1c}$, namely%
\begin{equation}
H_{1c}={\mathcal{O}}_{\Omega _{1}}\left( \mathrm{Ran}\,\Gamma \right) \text{
and, consequently, }H_{1d}=H_{1}\ominus H_{1c}.  \label{hhd3}
\end{equation}%
Based on this representation we deduce a condition that implies
the existence of decoupled states $H_{1d}$ in the presence of high symmetry in the internal dynamics in $H_1$ (corresponding to high multiplicity of $\Omega_1$):
\begin{equation}
\limfunc{mult}(\Omega _{1}\restrict H_{1c})\leq\limfunc{rank}\Gamma ,  \label{hhd4}
\end{equation}
where $\limfunc{mult}\left\{ \cdot \right\} $ and $\limfunc{rank}\left\{
\cdot \right\} $ are the spectral multiplicity and the rank of
an operator.  We prove this inequality later on in Theorem \ref{thmMultiplicity}.
If $\Omega _{1}$ and $\Gamma $ are generic, the inequality (\ref%
{hhd4}) would also be necessary for the existence of decoupled states. We
will refer to the condition (\ref{hhd4}) as the \emph{spectral multiplicity
condition}. This condition (\ref{hhd4}) readily implies that \emph{an open
system with low rank coupling and large spectral multiplicity must have
decoupled (frozen) states}.

Below we construct a couple of simple examples of Hamiltonian open systems
having decoupled degrees of freedom.  A detailed discussion with theorems on the coupled and decoupled parts of the state variables is presented in Section
\ref{sectionReconstructibility}.

\subsection{An oscillatory system with frozen degress of freedom}

Let us consider an open oscillatory Hamiltonian system $S_{1}$ described by
momentum and coordinate variables $\left\{ p,q\right\} $ with $p,q\in 
\mathbb{R}^{N}$, where $N$ is finite natural number. Hence, the Hilbert
space of observable variables here is $H_{1}=\mathbb{R}^{2N}$. We assume this open system to be a part of a larger Hamiltonian system for which the
complimentary system $S_{2}$ of hidden degrees of freedom is described by
variables $\left\{ \pi ,\varphi \right\} $ with $\pi ,\varphi \in G$, where $%
G$ is a real Hilbert space, and, hence, $H_{2}=G\oplus G$. We don't write it
explicitly, but rather presume that the system evolves according to the
Hamilton equations with the total Hamiltonian to be of the form%
\begin{equation}
\mathsf{H}\left( p,q;\pi ,\varphi \right) =\mathsf{h}_{1}\left( p,q\right) +%
\mathsf{h}_{2}\left( \pi ,\varphi \right) +\mathsf{h}_{\limfunc{int}}\left(
q,\varphi \right)  \label{fd1}
\end{equation}%
where $\mathsf{h}_{1}$ and $\mathsf{h}_{2}$ are correspondingly the internal
energies of systems $S_{1}$ and $S_{2}$, and $\mathsf{h}_{\limfunc{int}}$ is the interaction energy between $S_{1}$ and $S_{2}$. We assume $\mathsf{h}_{1}$
and $\mathsf{h}_{\limfunc{int}}$ to be of the form 
\begin{equation}
\mathsf{h}_{1}\left( p,q\right) =\frac{\left( p,p\right) }{2m}+\frac{\xi
\left( q,q\right) }{2},\ \mathsf{h}_{\limfunc{int}}\left( q,\varphi \right)
=\dsum\limits_{j=1}^{J}\left[ \left( q,\gamma _{1j}\right) -\left( \varphi, \gamma
_{2j}\right) \right] ^{2}  \label{fd2}
\end{equation}%
where $m$ and $\xi $ are postive constants, $1\leq J<N$, $\gamma _{1j}\in 
\mathbb{R}^{N}$ and $\gamma _{2j}\in G$. Evidently we can always choose an
orthonormal system of vectors $\left\{ \tilde{e}_{1},\ldots ,\tilde{e}%
_{N}\right\} $ in $\mathbb{R}^{N}$ so that%
\begin{equation}
E_{\gamma }=\limfunc{span}\left\{ \gamma _{11},\ldots ,\gamma _{1J}\right\} =%
\limfunc{span}\left\{ \tilde{e}_{N-J+1},\ldots ,\tilde{e}_{N}\right\} ,
\label{fd2a}
\end{equation}%
and introduce the corresponding new variables $\tilde{p},\tilde{q}\in \mathbb{%
R}^{N}$ by%
\begin{equation}
q=\dsum\limits_{s=1}^{N}q_{s}e_{s}=\dsum\limits_{s=1}^{N}\tilde{q}_{s}\tilde{%
e}_{s}\text{ where }\left\{ e_{1},\ldots ,e_{N}\right\} \text{ is the
standard basis in }\mathbb{R}^{N}.  \label{fd2b}
\end{equation}%
Next we introduce an orthogonal decomposition%
\begin{equation}
\tilde{p}=\tilde{p}^{\prime }\oplus \tilde{p}^{\prime \prime },\ \tilde{q}=%
\tilde{q}^{\prime }\oplus \tilde{q}^{\prime \prime }\text{, where }\tilde{p}%
^{\prime \prime },\tilde{q}^{\prime \prime }\in E_{\gamma }\text{ and }%
\tilde{p}^{\prime },\tilde{q}^{\prime }\in \mathbb{R}^{N}\ominus E_{\gamma },
\label{fd3}
\end{equation}%
and recast the energies in (\ref{fd2}) as follows%
\begin{gather}
\mathsf{h}_{1}\left( p,q\right) =\mathsf{h}_{1}^{\prime }\left( \tilde{p}%
^{\prime },\tilde{q}^{\prime }\right) +\mathsf{h}_{1}^{\prime \prime }\left( 
\tilde{p}^{\prime \prime },\tilde{q}^{\prime \prime }\right) ,\ \text{where}
\label{fd4} \\
\mathsf{h}_{1}^{\prime }\left( \tilde{p}^{\prime },\tilde{q}^{\prime
}\right) =\frac{\left( \tilde{p}^{\prime },\tilde{p}^{\prime }\right) }{2m}+%
\frac{\xi \left( \tilde{q}^{\prime },\tilde{q}^{\prime }\right) }{2},\ 
\mathsf{h}_{1}^{\prime \prime }\left( \tilde{p}^{\prime \prime },\tilde{q}%
^{\prime \prime }\right) =\frac{\left( \tilde{p}^{\prime \prime },\tilde{p}%
^{\prime \prime }\right) }{2m}+\frac{\xi \left( \tilde{q}^{\prime \prime },%
\tilde{q}^{\prime \prime }\right) }{2},  \notag \\
\mathsf{h}_{\limfunc{int}}\left( q,\varphi \right) =\dsum\limits_{j=1}^{J}%
\left[ \left( \tilde{q}^{\prime \prime},\tilde{\gamma}_{1j} \right) 
-\left(\varphi,\gamma _{2j} \right) \right] ^{2}.  \notag
\end{gather}%
It is evident from (\ref{fd4}) that the variables $\left\{ \tilde{p}^{\prime
},\tilde{q}^{\prime }\right\} $ are decoupled from the system $S_{2}$, and
all the coupling from $S_{1}$ to $S_{2}$ is only through the variables $%
\left\{ \tilde{p}^{\prime \prime },\tilde{q}^{\prime \prime }\right\} $.  In fact, $S_1$ couples directly to $S_2$ through the variables $\tilde q''$ only, but this coupling affects $\tilde p''$ through the internal dynamics in $S_1$.  $\{ \tilde p', \tilde q' \}$ remain, however, unaffected.  This together with (\ref{fd3}) yields the following estimates for the space $H_{1d}$ of ``decoupled" states $\{ \tilde p', \tilde q' \}$.
\begin{equation}
H_{1d}\supseteq \left( \mathbb{R}^{N}\ominus E_{\gamma }\right)^2
\text{, and, hence, }\dim
H_{1d}\geq 2\left( N-J\right).  \label{fd4a}
\end{equation}
An elementary analysis of the used arguments shows that the existence of
decoupled variables in the above example is due to (i) the highly symmetric
form of the Hamiltonaian $\mathsf{h}_{1}\left( p,q\right) $ in (\ref{fd2}),
resulting in the maximal spectral multiplicity $N$, and (ii) the coupling
of rank $J$, which is less than $N$ and application of the spectral multiplicity condition \eqref{hhd4}.

Notice that, if instead of (\ref{fd2}), we would have
\begin{equation}
\mathsf{h}_{1}\left( p,q\right) =\dsum\limits_{s=1}^{N}\frac{p_{s}^{2}}{2m_{s}}+\dsum\limits_{s=1}^{N}\frac{\xi _{s}q_{s}^{2}}{2}  \label{fd5}
\end{equation}
with all different and generic $m_{s}$ and $\xi _{s}$, then the
corresponding spectral multiplicity would be one and there will be no
decoupled degrees of freedom.
We point out also that, in this case, for a generic $\gamma _{1j}$ in the
representation (\ref{fd2}) every vector from the original orthonormal system 
$e_{1},\ldots ,e_{N}$ in $\mathbb{R}^{N}$ has nonzero projections onto
both $E_{\gamma }$ and $\mathbb{R}^{N}\ominus E_{\gamma }$, implying that 
\emph{generically none of the original variables }$\left\{
p_{s},q_{s}\right\} $\emph{\ can be considered as being decoupled from the
system }$S_{2}$. This indicates that decoupling of variables due the
spectral multiplicity, though elementary, is not trivial.

\subsection{Toy model of a solid with frozen degrees of freedom}

We construct here a toy model for a solid having frozen degrees of freedom
due to high spectral multiplicity, naturally arising from system symmetries.
Let us consider the $d$-dimensional lattice
\begin{equation}
\mathbb{Z}^{d}=\left\{ n:n=\left( n_{1},\ldots ,n_{d}\right) ,\ n_{j}\in 
\mathbb{Z}\right\} \text{ where }\mathbb{Z}\text{ is the set of integers,}
\label{tms1}
\end{equation}
and introduce a system $S$ as a lattice array of identical oscillatory
systems similar to that described in the previous section. Namely, we assume
that the system state is of the form
$u=\left\{ \left[ p_{n},q_{n}\right] ,\ n\in \mathbb{Z}^{d}\right\}$
where with $p_{n},q_{n}\in \mathbb{R}^{N}$, where $N$ is a finite natural number.

The system Hamiltonian $\mathsf{H}\left( p,q \right) $ is assumed to be spatially homogeneous, local, and of the form
\begin{equation}
\mathsf{H}\left( p,q\right) =\dsum\limits_{n\in \mathbb{Z}^{d}}\left[ 
\mathsf{h}_{1}\left( p_{n},q_{n}\right) +\dsum\limits_{j=1}^{J}\| (\nabla
q_{n},\gamma _{j}) \|^{2}\right] ,\ p,q\in H,  \label{tms2}
\end{equation}
where the local Hamiltonian $\mathsf{h}_{1}\left( p,q\right) $ is defined by
(\ref{fd2}), and the vectors $\gamma _{j}\in \mathbb{R}^{N}$, $1\leq j\leq J$, describe the interactions between neighboring sites through the discrete gradient $\nabla$.
An expansion of the inner sum gives
\begin{equation}  \label{tms2b}
\sum_{j=1}^J \| \nabla_n (q_n,\gamma_j) \|^2 =
\sum_{j=1}^J\sum_{i=1}^d
\left( (q_{n},\gamma_j) - (q_{n+e_i},\gamma_j) \right)^2,
\end{equation}
in which $e_i = (\delta_{i1},\dots,\delta_{in})$.
Now denoting
\begin{equation}
\left\vert m\right\vert _{0}=\max_{1\leq j\leq d}\left\vert m_{j}\right\vert
,\ m=\left( m_{1},\ldots ,m_{d}\right) \in \mathbb{Z}^{d}  \label{tms3}
\end{equation}
we consider an arbitrary finite lattice cube
\begin{equation}
\Lambda =\Lambda _{L}=\left\{ n\in \mathbb{Z}^{d}:\left\vert n\right\vert
_{0}\leq L\right\} \text{ where }L\geq 2\text{ is an integer,}  \label{tms4}
\end{equation}
and define its volume $\left\vert \Lambda \right\vert $ by
\begin{equation}
\left\vert \Lambda \right\vert =\text{number of sites } n\in \Lambda .  \label{tms5}
\end{equation}
Now we introduce a system $S_{\Lambda }$ associated with the finite lattice cube $\lambda$, in which the states are functions from $\Lambda$ to
$\mathbb{R}^N \oplus \mathbb{R}^N$, or
$\left\{ \left[ p_{n},q_{n}\right] ,\ n\in \Lambda \right\}$, with the Hamiltonian
\begin{equation}
\mathsf{H}_{\Lambda }\left( p,q\right) =\dsum\limits_{n\in \Lambda }\left[ 
\mathsf{h}_{1}\left( p_{n},q_{n}\right) +\dsum\limits_{j=1}^{J_{0}}
\|\nabla q_{n},\gamma _{j}\|^{2}\right] ;\ q_{n}=0 \text{ for } n\notin \Lambda .
\label{tms6}
\end{equation}
Recall now that the system dynamics is described then by the Hamilton
equations
\begin{equation}
\frac{dp}{dt}=-\frac{dH}{dq},\ \frac{dq}{dt}=\frac{dH}{dp},  \label{pht1}
\end{equation}
which, in our case, turns into the linear evolution equation of the form
\begin{equation}
\frac{du}{dt}=-\mathrm{i}\Omega u,\ u=\left[ p,q\right]  \label{pht2}
\end{equation}
in which multiplication by $i$ is defined by $i[p,q] = [-q,p]$.
Without writing the relevant operator (matrix) $\Omega$ explicitly, we simply
denote by $\Omega_\Lambda$ the respective matrix for the Hamiltonian
$\mathsf{H}_{\Lambda }$.  Observe now that in view of the form (\ref{tms6}) of
the Hamiltonian $\mathsf{H}_{\Lambda}$, an open oscillatory system
associated with any single site $n\in \Lambda$ is exactly of the form
considered in the previous section (see \eqref{tms2} and \eqref{tms2b}),
and, consequently, it has decoupled degrees of freedom described by the
space $\left( \mathbb{R}^{N}\ominus E_{\gamma }\right) ^{2}$ not depending
on $n$.  This implies that the space of decoupled (frozen) states $H_{d}$ satisfies
\begin{equation}
H_{d}\supseteq \left( \mathbb{R}^{N}\ominus E_{\gamma }\right) ^{2|\Lambda|}.
\label{pht2a}
\end{equation}
If we put $H = H_{1d} \oplus H_{1c}$, then this, combined with the fact the local Hamiltonians ($\mathsf{h}_1$) at all sites are identical gives the rank of the coupling as $J|\Lambda|$.   By the
spectral multiplicity condition \eqref{hhd4} we then obtain a bound on the spectral multiplicity of the restriction $\Omega_\Lambda\restrict H_c$ to the ``coupled" part $H_c$:
\begin{equation}
\limfunc{mult}(\Omega _{\Lambda }\restrict H_c) \leq J \left\vert \Lambda
\right\vert =
J \left( 2L+1\right)^{d}.  \label{pht3}
\end{equation}
In fact, a more eleborate analysis based on introduction of lattice toruses
along with lattice cubes can produce an approximate formlula of the
following form
\begin{equation}
\frac{\limfunc{mult}\Omega _{\Lambda }}{\left\vert \Lambda \right\vert }%
=C_{0}+O\left( \left\vert \Lambda \right\vert ^{-\frac{1}{d}}\right) ,
\label{pht4}
\end{equation}
where $C_{0}$ is a constant similar to $J$.

The estimates (\ref{pht3}) and (\ref{pht4}) for the solid toy model indicate
that the high spectral multiplicity can cause many degrees of freedom to be
comletilely decoupled from the rest of the system.

\section{Reconstructibility from open subsystems}

\label{sectionReconstructibility}

According to our general strategy, we regard an open system within its
conservative system from two points of view that are closely related. In the
first, we consider a conservative system composed of two coupled subsystems, each
treated equally. Because of the coupling, the subsystems are open, and we
analyze the extent to which the conservative system is reconstructible from either
of its open subsystems. In the second, the objects that play the leading
role are the ``master" conservative system and a given subsystem. The subsystem is
open, as it interchanges energy with the master system.

\subsection{Two coupled open systems}

\label{subsecReconstructiblilty1}

We first investigate a conservative system composed of two coupled open ones.
Typically, one system will be observable, such as a resonator, and the other
will represent its ``exterior" which we associate with the hidden degrees of
freedom.

Let us begin with two conservative systems, Closed System 1, identified by
the triple $(H_{1},m_{1},A_{1})$ that represents an observable system
\begin{equation}
m_{1}\partial _{t}v_{1}(t)=-\mathrm{i}A_{1}v_{1}(t)+f_{1}(t)\quad \text{in }
H_{1}
\end{equation}
satisfying the rest condition (Condition \ref{condRest}, page \pageref{condRest}), and another linear system, Closed System 2, identified by the triple
$(H_{2},m_{2},\Omega _{2})$ that represents the system of hidden variables
\begin{equation}
m_{2}\partial _{t}v_{2}(t)=-\mathrm{i}A_{2}v_{2}(t)+f_{2}(t)\quad \text{in }%
H_{2}
\end{equation}
also satisfying the rest condition.
$A_{1}$ and $A_{2}$ are self-adjoint, and the mass operators $m_{1}$ and
$m_{2}$ are positive. We then couple the two systems through a bounded
operator $\Gamma $ and its adjoint: 
\begin{equation}
\Gamma :H_{2}\rightarrow H_{1}\text{,}\qquad \Gamma ^{\dagger
}:H_{1}\rightarrow H_{2}.
\end{equation}%
The conservative system as composed of these two subsystems then has the form 
\begin{align}
m_{1}\partial _{t}v_{1}\left( t\right) & =-\mathrm{i}A_{1}v_{1}\left(
t\right) -\mathrm{i}\Gamma v_{2}\left( t\right) +f_{1}\left( t\right) ,
\label{sm} \\
m_{2}\partial _{t}v_{2}\left( t\right) & =-\mathrm{i}\Gamma ^{\dagger
}v_{1}\left( t\right) -\mathrm{i}A_{2}v_{2}\left( t\right) +f_{2}\left(
t\right) .  \notag
\end{align}%
Using the following rescaling transformation 
\begin{equation}
v_{j}\rightarrow m_{j}^{-\frac{1}{2}}v_{j},\quad A_{j}\rightarrow m_{j}^{%
\hspace*{0.5ex}\frac{1}{2}}\Omega _{j}m_{j}^{\hspace*{0.5ex}\frac{1}{2}%
},\quad f_{j}\rightarrow m_{j}^{\hspace*{0.5ex}\frac{1}{2}}f_{j},\quad
\Gamma \rightarrow m_{1}^{\hspace*{0.5ex}\frac{1}{2}}\Gamma m_{2}^{\hspace*{%
0.5ex}\frac{1}{2}}
\end{equation}%
we recast the system (\ref{sm}) into the simpler form 
\begin{align}
\partial _{t}v_{1}\left( t\right) & =-\mathrm{i}\Omega _{1}v_{1}\left(
t\right) -\mathrm{i}\Gamma v_{2}\left( t\right) +f_{1}\left( t\right) ,
\label{smb} \\
\partial _{t}v_{2}\left( t\right) & =-\mathrm{i}\Gamma ^{\dagger
}v_{1}\left( t\right) -\mathrm{i}\Omega _{2}v_{2}\left( t\right)
+f_{2}\left( t\right) ,  \notag
\end{align}%
which we will use from now on. We refer to the operators $\Omega _{1}$ and $%
\Omega _{2}$ as the \emph{frequency operators} for the observable and hidden
systems.

In matrix form, the system (\ref{smb}) is written as 
\begin{equation}
\partial _{t}\mathcal{V}\,=\,-\mathrm{i}\Omega \,\mathcal{V}+\mathcal{F},
\quad \mathcal{V},\mathcal{F}\,\in \mathcal{H}=H_{1}\oplus H_{2},
\label{smb1}
\end{equation}%
in which the frequency operator $\Omega $ has the block-matrix structure 
\begin{equation}  \label{Omega}
\Omega =\left[ 
\begin{array}{cc}
\Omega _{1} & \Gamma \\ 
\Gamma ^{\dagger } & \Omega _{2}%
\end{array}%
\right] .
\end{equation}
Because of the coupling, both systems become open. Their dynamics are
obtained by projecting the dynamics of the large conservative system in $\mathcal{H%
}$ to $H_{1}$ and $H_{2}$ separately. For System~1, this means setting the
forcing from the second equation of the system (\ref{smb}) to zero ($%
f_{2}(t)=0$), solving for $v_{2}$, and then inserting the result into the
first equation. This, together with an analogous computation for System~2,
results in the dynamical equations for the open systems Open System 1 $%
(H_{1},\Omega _{1},a_{1}(t))$ and Open System 2 $(H_{2},\Omega
_{2},a_{2}(t)) $: 
\begin{align}
\partial _{t}v_{1}(t)& =-\mathrm{i}\Omega _{1}v_{1}(t)-\int_{0}^{\infty
}a_{1}(\tau )v_{1}(t-\tau )\,\mathrm{d}\tau +f_{1}(t)\quad \text{in }H_{1},
\label{reduced1} \\
\partial _{t}v_{2}(t)& =-\mathrm{i}\Omega _{2}v_{2}(t)-\int_{0}^{\infty
}a_{2}(\tau )v_{2}(t-\tau )\,\mathrm{d}\tau +f_{2}(t)\quad \text{in }H_{2},
\label{reduced2}
\end{align}%
in which 
\begin{equation}
a_{1}(t)=\Gamma \mathrm{e}^{-\mathrm{i}\Omega _{2}t}\Gamma ^{\dagger }\quad 
\text{and}\quad a_{2}(t)=\Gamma ^{\dagger }\mathrm{e}^{-\mathrm{i}\Omega
_{1}t}\Gamma ,
\end{equation}%
and the functions $a_{j}(t)$ are the friction functions. The rest condition, Condition \ref{condRest} continues to hold, and, by virtue of their form, the equations automatically satisfy the power dissipation condition, Condition \ref{condDissipation}.

\emph{We ask the question: How much of }$H_{2}$\emph{\ can be reconstructed
from Open System 1 (\ref{reduced1}) alone; in other words, how much
information about the hidden variables is encoded in the friction function }
$a_{1}(t)$\emph{\ for the observable variables}? We can view $a_{1}(t)$ as a
dynamical mechanism by which an observer confined to the observable state
variables detects or influences the hidden degrees of freedom. 
The subspace of $H_{2}$ that is reconstructible by $a_{1}(t)$ we call the 
\emph{coupled component of $H_{2}$} and denote it by $H_{2c}$. Clearly this
subspace is determined by the coupling channels to $H_{2}$ given by $\Gamma
^{\dagger }$ and the internal action by $\Omega _{2}$ on $H_{2}$; this is
explicitly evident in the form $a_{1}(t) = \Gamma \mathrm{e}^{-\mathrm{i}
\Omega _{2}t}\Gamma ^{\dagger }$.

The question of the extent to which the observable system determines the
hidden is tantamount to that of \emph{determining the unique minimal
conservative extension of Open System 1 within the large system
$(\mathcal{H},\Omega )$} (see Section \ref{secExtension}). According to Proposition \ref{pmin1} the Hilbert state space $\mathcal H_{\min }\supset H_{1}$ of this conservative
extension is simply the orbit (see Definition \ref{dorbit})
of $H_{1}$ under the action of $\Omega $, as a subspace of $\mathcal H$:
\begin{equation}
\mathcal H_{\min }={\mathcal{O}}_{\Omega }(H_{1})=H_{1}\oplus H_{2c}, \quad
H_{2c}:=\mathcal H_{\min }\ominus H_{1}.  \label{smb3}
\end{equation}%
In addition, it can be shown that $H_{2c}\subseteq H_{2}$ is invariant under
the action of $\Omega _{2}$, namely 
\begin{equation}
\Omega _{2}H_{2c}\subseteq H_{2c}\subseteq H_{2},  \label{smb4}
\end{equation}%
and we refer to $H_{2c}$ as the \emph{coupled component of $H_{2}$}. This
construction of $H_{2c}$ $\subseteq H_{2}$ gives rise to a canonical
$\Omega_2$-invariant orthogonal decomposition of the hidden state variables, 
\begin{equation}
H_{2}=H_{2c}\oplus H_{2d}.
\label{decomposition2}
\end{equation}%
We refer to the subspace $H_{2d}$ defined in (\ref{decomposition2}) as
the \emph{decoupled component of $H_{2}$}.  The systems in $H_{2c}$ and
$H_{2d}$ evolve independently in time by the internal dynamics $\Omega _{2}$ of the hidden variables $H_{2} $. Furthermore, since it can be shown that the operator range $\mathrm{Ran}(\Gamma ^{\dagger })\subset H_{2c}$, we see that no forcing
function from $H_{2d}$ can influence the dynamics of the observable
variables and, in turn, does not influence the dynamics of $H_{2c}$ through
the coupling.

In an analogous way, we can ask, how much of $H_{1}$ can a hypothetical
observer confined to $H_{2}$ reconstruct? The component $H_{1c}$ of $H_{1}$
that is reconstructible by an observer in the hidden state variables we call
the \emph{coupled component of $H_{1}$}. Its orthogonal complement $H_{1d}$
is the \emph{decoupled component of $H_{1}$}, and we have the decomposition 
\begin{equation}
H_{1}=H_{1c}\oplus H_{1d},  \label{decomposition1}
\end{equation}
which is invariant under the action of $\Omega _{1}$. The state space of the
unique minimal conservative extension of Open System 2 is $H_{1c}\oplus
H_{2} $.

With respect to these decompositions of the observable and hidden variables
into the coupled and decoupled components $\mathcal{H}=H_{1d}\oplus
H_{1c}\oplus H_{2c}\oplus H_{2d}$, the frequency operator $\Omega $ for the
conservative system \eqref{smb} has the matrix form 
\begin{equation}
\Omega =\left[ 
\begin{array}{cccc}
\Omega _{1d} & 0 & 0 & 0 \\ 
0 & \Omega _{1c} & \Gamma _{c} & 0 \\ 
0 & \Gamma _{c}^{\dagger } & \Omega _{2c} & 0 \\ 
0 & 0 & 0 & \Omega _{2d}%
\end{array}%
\right] ,  \label{omg1}
\end{equation}%
in which the subscripts refer to retrictions of the domain: 
\begin{equation}
\Omega _{ic}=\Omega \!\!\upharpoonright \!\!H_{ic},\quad \Omega
_{id}=\Omega \!\!\upharpoonright \!\!H_{id},\quad {i=1,2},\qquad \text{and}%
\quad \Gamma _{c}=\Gamma \!\!\upharpoonright \!\!H_{2c}.  \label{omg2}
\end{equation}%
We can see from (\ref{omg1}) that the decoupled parts $H_{1d}$ and $H_{2d}$,
can be analyzed independently of the rest of the system justifying their
name ``decoupled". Furthermore, the conservative subsystem $(H_{1c}\oplus
H_{2c},\Omega _{c})$ with frequency operator 
\begin{equation}
\Omega _{c}=\left[ 
\begin{array}{cc}
\Omega _{1c} & \Gamma _{c} \\ 
\Gamma _{c}^{\dagger } & \Omega _{2c}%
\end{array}%
\right] ,  \label{omg3}
\end{equation}%
which consists of the part of $H_{1}$ reconstructible by Open System 2 alone
and the part $H_{2}$ reconstructible by Open System 1 alone, is itself fully
reconstructible by either of the open subsystems $(H_{1c},\Omega
_{1c},a_{1}(t))$ or $(H_{2c},\Omega _{2c},a_{2}(t))$. This is equivalent to
the statement that {\em $(H_{1c}\oplus H_{2c},\Omega _{c})$ is the unique minimal
conservative extension, realized as a subsystem of $(\mathcal{H},\Omega )$,
of each of its open components separately.} This motivates the following
definition.

\begin{definition}[reconstructibility]
\label{reconstructible} A conservative linear system composed of two coupled
subsystems 
\begin{align}
\partial _{t}v_{1}\left( t\right) & =-\mathrm{i}\Omega _{1}v_{1}\left(
t\right) -\mathrm{i}\Gamma v_{2}\left( t\right) + f_1(t) ,  \notag \\
\partial _{t}v_{2}\left( t\right) & =-\mathrm{i}\Gamma ^{\dagger
}v_{1}\left( t\right) -\mathrm{i}\Omega _{2}v_{2}\left( t\right) + f_2(t),  \notag
\end{align}%
with $v_{1}(t)\in H_{1}$ and $v_{2}(t)\in H_{2}$ is called \emph{%
reconstructible} if it is the minimal conservative extension of each of the
open projected linear systems 
\begin{align}
\partial _{t}v_{1}(t)& =-\mathrm{i}\Omega _{1}v_{1}(t)-\int_{0}^{\infty
}\Gamma \mathrm{e}^{-\mathrm{i}\Omega _{2}\tau }\Gamma ^{\dagger
}\,v_{1}(t-\tau )\,\,\mathrm{d}\tau  + f_1(t) \qquad \text{in }H_{1} , \notag \\
\partial _{t}v_{2}(t)& =-\mathrm{i}\Omega _{2}v_{2}(t)-\int_{0}^{\infty
}\Gamma ^{\dagger }\mathrm{e}^{-\mathrm{i}\Omega _{1}\tau }\Gamma
\,v_{2}(t-\tau )\,\,\mathrm{d}\tau + f_2(t) \qquad \text{in }H_{2} .  \notag
\end{align}%
In other words, the conservative system $(H_1\oplus H_2,\Omega)$ ($\Omega$ is
defined by its decomposition \eqref{Omega}) is reconstructible if all of $%
H_{2} $ can be reconstructed from the open system $(H_{1},\Omega
_{1},a_{1}(t))$ and all of $H_{1}$ can be reconstructed from the open system 
$(H_{2},\Omega _{2},a_{2}(t))$.

We say that $H_2$ is \emph{reconstructible from} the open system $%
(H_1,\Omega_1,a_1(t))$ if $H_1\oplus H_2$ is (isomorphic to) the state space for the minimal conservative extension of $(H_1,\Omega_1,a_1(t))$.
\end{definition}

A conservative system that is reconstructible may possibly be further decomposed
into independent conservative subsystems that commute with the projection to $%
H_{1} $, in other words, that are of the form $H_{1}^{\prime }\oplus
H_{2}^{\prime }$ with $H_{1}^{\prime }\subseteq H_{1}$ and $H_{2}^{\prime
}\subseteq H_{2}$. However, $H_{1}^{\prime }$ will not contain the entire
range of $\Gamma $ and accordingly will not contain all of the information
of the delayed response function $a_{1}(t)$. We take up these finer
decompositions further in Section \ref{sectionDecomposition}.

The simplest reconstructibility theorem is as follows. More general
reconstructibility statements as well as the proofs are given in Section \ref%
{sectionProofs}.

\begin{theorem}[system reconstructibility]
\label{thmReconstructibility} Let a conservative "master" system composed of two coupled systems be given: 
\begin{align}
\partial _{t}v_{1}\left( t\right) & =-\mathrm{i}\Omega _{1}v_{1}\left(
t\right) -\mathrm{i}\Gamma v_{2}\left(t\right) + f_1(t) , \\
\partial _{t}v_{2}\left( t\right) & =-\mathrm{i}\Gamma ^{\dagger
}v_{1}\left( t\right) -\mathrm{i}\Omega _{2}v_{2}\left(t\right) + f_2(t),
\end{align}%
with $v_{1}(t)$ in the state space $H_{1}$ and $v_{2}(t)$ in the state space 
$H_{2}$, and let the coupling operator $\Gamma :H_{2}\rightarrow H_{1}$ be
bounded. \ Let also $H_{2c}$ denote the subspace of $H_{2}$ that is
reconstructible from Open System 1 and $H_{1c}$ the subspace of $H_{1}$ that
is reconstructible from Open System 2.

\begin{enumerate}
\item $H_{1c}$ consists of the set of states of $H_1$ that are accessible by
applying the internal dynamics of $H_1$ (given by $\Omega_1$) to all vectors
in $H_1$ to which $H_2$ is directly coupled by $\Gamma$, that is, 
\begin{equation}
H_{1c} = {\mathcal{O}}_{\Omega_1}(\mathrm{Ran}\,\Gamma).
\end{equation}
Similarly, 
\begin{equation}
H_{2c} = {\mathcal{O}}_{\Omega_2}(\mathrm{Ran}\,\Gamma^\dagger).
\end{equation}

\item The restriction of the master system to $H_{1}\oplus H_{2c}$ is the unique minimal conservative extension of $H_1$, and the restriction to $H_{1c}\oplus H_{2}$ is the unique minimal conservative extension of $H_2$.

\item The restriction of the master system to $H_{1c}\oplus H_{2c}$ is the unique reconstructible subsystem of the master system that completely determines the friction
functions of Open Systems 1 and 2, namely $\Gamma \mathrm{e}^{-\mathrm{i}%
\Omega _{2}t}\Gamma ^{\dagger }$ and $\Gamma ^{\dagger }\mathrm{e}^{-\mathrm{%
i}\Omega _{1}t}\Gamma $.

\item $(H_{1}\!\oplus\! H_{2},\Omega)$ is reconstructible if and only if $%
H_1 $ and $H_2$ have no nontrivial $\Omega$-invariant subspaces.
\end{enumerate}
\end{theorem}

{\bfseries Example: two coupled finite systems.} \
With a simple finite-dimensional example of two coupled open systems, we
illustrate the interaction between the two components and the extent to
which each is determined, or reconstructible, by the other. The observations
are generalized and proved in Theorem \ref{thmReconstructibility}.

Let us begin with the state space of observable variables $H_{1}=\mathbb{C}%
^{2}$, with variable vector $v\in H_{1}$ and an open DD system 
\begin{equation}
\partial _{t}v(t)=-\mathrm{i}\left[ 
\begin{array}{cc}
a\! & \!b \\ 
b^{\ast }\! & \!c%
\end{array}%
\right] v(t)-\mathrm{i}\left[ 
\begin{array}{c}
\alpha \\ 
\beta%
\end{array}%
\right] \left[ 
\begin{array}{cc}
\alpha ^{\ast } & \beta ^{\ast }%
\end{array}%
\right] \int_{0}^{\infty }\left( |\gamma |^{2}\mathrm{e}^{-\mathrm{i}\mu
_{1}\tau }+|\delta |^{2}\mathrm{e}^{-\mathrm{i}\mu _{2}\tau }\right)
v(t-\tau )\,\mathrm{d}\tau ,
\end{equation}%
in which we assume $|\alpha |^{2}+|\beta |^{2}=1$ and $\mu _{1}$ and $\mu
_{2}$ are real. The operator for the internal dynamics in $H_{1}$ is 
\begin{equation}
\Omega _{1}=\left[ 
\begin{array}{cc}
a\! & \!b \\ 
b^{\ast }\! & \!c%
\end{array}%
\right] =\left[ 
\begin{array}{cc}
\alpha \! & \!-\beta ^{\ast } \\ 
\beta \! & \!\alpha ^{\ast }%
\end{array}%
\right] \left[ 
\begin{array}{cc}
\lambda _{1}\! & 0 \\ 
0\! & \!\lambda _{2}%
\end{array}%
\right] \left[ 
\begin{array}{cc}
\alpha ^{\ast }\! & \!\beta ^{\ast } \\ 
-\beta \! & \!\alpha%
\end{array}%
\right] ,  \label{Omega1}
\end{equation}%
in which $a$, $c$, $\lambda _{1}$, and $\lambda _{2}$ are real. The
delayed-response function 
\begin{equation}
\mathrm{i}a_{1}(t)=\left[ 
\begin{array}{c}
\alpha \\ 
\beta%
\end{array}%
\right] \left[ 
\begin{array}{cc}
\alpha ^{\ast } & \beta ^{\ast }%
\end{array}%
\right] \left( |\gamma |^{2}\mathrm{e}^{-\mathrm{i}\mu _{1}\tau }+|\delta
|^{2}\mathrm{e}^{-\mathrm{i}\mu _{2}\tau }\right)
\end{equation}%
involves two frequencies, each with a matrix factor of rank one. The space $%
H_{2}$ of hidden variables is therefore isomorphic to $\mathbb{C}^{2}$; in
fact, 
\begin{equation}
\mathrm{i}a_{1}(t)=\Gamma \mathrm{e}^{-\mathrm{i}\Omega _{2}t}\Gamma
^{\dagger },
\end{equation}%
where 
\begin{equation}
\Gamma =\left[ 
\begin{array}{c}
\alpha \\ 
\beta%
\end{array}%
\right] \left[ 
\begin{array}{cc}
\gamma ^{\ast } & \delta ^{\ast }%
\end{array}%
\right] ,\ \Omega _{2}=\left[ 
\begin{array}{cc}
\mu _{1} & 0 \\ 
0 & \mu _{2}%
\end{array}%
\right] .
\end{equation}%
The minimal conservative extension of $(H_{1},\Omega _{1},a_{1}(t))$ is $(%
\mathcal{H},\Omega )$, where 
\begin{equation}
\mathcal{H}=H_{1}\oplus H_{2},\ \Omega =\left[ 
\begin{array}{cc|cc}
a & b & \alpha \gamma ^{\ast } & \alpha \delta ^{\ast } \\ 
b^{\ast } & c & \beta \gamma ^{\ast } & \beta \delta ^{\ast } \\ \hline
\alpha ^{\ast }\gamma & \beta ^{\ast }\gamma & \mu _{1} & 0 \\ 
\alpha ^{\ast }\delta & \beta ^{\ast }\delta & 0 & \mu _{2}%
\end{array}%
\right] .
\end{equation}%
In this particular example, $\Gamma $ has rank $1$ because the matrices in $%
\mathrm{i}a_{1}(t)$ for the two frequencies have the same range. This range,
which is the range of $\Gamma $, happens to be an eigenspace for $\Omega
_{1} $ corresponding to the eigenvalue $\lambda _{1}$ (equation \ref{Omega1}%
). The delayed-response function for the reduced dynamics in $H_{2}$
therefore only involves this single frequency: 
\begin{equation}
a_{2}(t)=\left[ 
\begin{array}{c}
\gamma \\ 
\delta%
\end{array}%
\right] \left[ 
\begin{array}{cc}
\gamma ^{\ast } & \delta ^{\ast }%
\end{array}%
\right] \mathrm{e}^{\mathrm{i}\lambda _{1}t}.
\end{equation}%
The state space of the minimal conservative extension of $(H_{2},\Omega
_{2},a_{2}(t))$ is the three-dimensional space 
\begin{equation}
H_{1c}\oplus H_{2},
\end{equation}%
in which $H_{1c}$ is the \textquotedblleft coupled component" of $H_{1}$,
consisting of the eigenspace for the eigenvalue $\lambda _{1}$.

If one projects the dynamics to $H_{1c}$, then its minimal conservative
extension is the same as that of $(H_2,\Omega_2,a_2(t))$. Thus, the space $%
H_2$, as well as its internal dynamics operator $\Omega_2$ and the coupling $%
\Gamma$, are reconstructible from the dynamics projected to $H_{1c}$, just
as $H_{1c}$, $\Omega_1$ restricted to $H_{1c}$, and $\Gamma$ are
reconstructible from $(H_2,\Omega_2,a_2(t))$. We therefore call the system
in $H_{1c} \oplus H_2$ \emph{reconstructible} (Definition \ref%
{reconstructible}).

In Theorem \ref{thmReconstructibility}, we prove that a coupled pair of open
systems forming a conservative system, $H_1\oplus H_2$ admits a unique
reconstructible subsystem system $H_{1c}\oplus H_{2c}$, containing all the
information of $a_1(t)$, in which the projection of the dynamics to each
part is sufficient to reconstruct the other. As in the simple example of
this subsection, it is always true that $H_{1c}$ is the $\Omega_1$-orbit of $%
\mathrm{Ran}\,\Gamma$ and $H_{2c}$ is the $\Omega_2$-orbit of $\mathrm{Ran}%
\,\Gamma^\dagger$.

\subsection{Open subsystems of conservative systems}

\label{subsecReconstructiblilty2}

Often an open system arises as a part of a given conservative system $(\mathcal H,\Omega)$ projected onto an observable subspace $H_1\subset \mathcal H$.
We investigate the way in which the state space
$\mathcal{H}_\text{{\scriptsize {min}}}$ of the minimal conservative
extension of the open system in $H_{1}$ is reconstructed within the spectral
structure of $(\mathcal{H},\Omega )$. We shall see that $\mathcal{H}_\text{%
{\scriptsize {min}}}$ is generated by \emph{the projections of all vectors
in $H_1$ onto the eigenspaces of $\Omega$}, as well as by \emph{the
projections, onto the eigenspaces of $\Omega$, of those vectors in $H_1$ and $%
\mathcal{H}\ominus H_1$ that are directly coupled through $\Gamma$} (the
ranges of $\Gamma$ and $\Gamma^\dag$). We discuss both points of view.
We investigate similar constructions for the generation of $H_{1c}$ and $H_{2c}$ by eigenmodes of $\Omega_1$ and $\Omega_2$ and arrive at one of our main results, Theorem \ref{thmMultiplicity}, which bounds the number of extending strings by the rank of the coupling.

In the case, say, of a finite resonator embedded within an infinite planar
lattice, for which the multiplicity of each eigenvalue is infinite, this
result has immediate consequences: namely, the multiplicities of the
eigenfrequencies for the minimal extension $(\mathcal{H}_\text{{\scriptsize {%
min}}}, \Omega\!\!\upharpoonright\!\!\mathcal{H}_\text{{\scriptsize {min}}}%
)$ are uniformly bounded and hence $\mathcal{H}_\text{{\scriptsize {min}}}$
is but a very small part of $\mathcal{H}$. The perhaps more interesting
point of view, in which the directly coupled modes (the ranges of $\Gamma$
and $\Gamma^\dagger$) generate $\mathcal{H}_\text{{\scriptsize {min}}}$, has
special significance for an object in surface contact with an infinite
medium.

\medskip

\noindent {\bfseries \ref{subsecReconstructiblilty2}.1 Generating the
minimal extension from the observable states.} \ Not surprisingly, \emph{all
of the modes (eigenfunctions) of the frequency operator $\Omega$ that
contribute to the $\Omega$-mode decomposition of any one of the vectors in the
``observable" space $H_1$ must be included as states of the minimal
extension.} These modes, in turn, generate all of the observable vectors,
and therefore the entire space $\mathcal{H}_\text{{\scriptsize {min}}}$.

To understand this, let us begin with the case in which $\mathcal{H}$ is
finite dimensional. Let $\{\lambda_\alpha\}_{\alpha=1}^n$ be the distinct
eigenvalues of $\Omega$. We then have a decomposition of $\mathcal{H}$ into
orthogonal eigenspaces 
\begin{equation}
\mathcal{H } = \bigoplus\limits_{\alpha=1}^n \mathcal{H}^\alpha \,,
\end{equation}
with respect to which the operator $\Omega$ is diagonal: 
\begin{equation}
\text{for each} \quad v=\sum_{\alpha=1}^n v_\alpha, \quad \Omega v =
\sum_{\alpha=1}^n \lambda_\alpha v_\alpha.
\end{equation}
As we have discussed, $\mathcal{H}_\text{{\scriptsize {min}}}$ is the
subspace of $\mathcal{H}$ that is generated by the vectors in $H_{1}$
through the operator $\Omega $, in other words, it is the orbit
${\mathcal{O}}_{\Omega }(H_{1})$ of $H_{1}$ under $\Omega $. In the finite-dimensional case, this is simply the vector space spanned by the vectors $\Omega^k(v)$ for $v\in H_1$ and $0\leq k<n$. Since the monomials $\lambda^k$,
restricted to the spectrum of $\Omega$, span the space of functions defined
on the spectrum, we have $v_\alpha\in \mathcal{H}_\text{{\scriptsize {min}}}$
for each $\alpha=1,\dots, n$. Now since the vector $v\in H_1$ is in turn
generated through linear combination by its projections $v_\alpha$, we
conclude that $\mathcal{H}_\text{{\scriptsize {min}}}$ is generated by the
projections of $H_1$ onto the eigenspaces of $\Omega$, denoted by $%
\pi_\alpha(H_1)$: 
\begin{equation}  \label{HminSpectrum1a}
\mathcal{H}_\text{{\scriptsize {min}}} = {\mathcal{O}}_{\Omega}(H_1) =
\bigoplus\limits_{\alpha=1}^n \pi_\alpha(H_1).
\end{equation}
Thus we have a spectral decomposition for $(\mathcal{H}_\text{{\scriptsize {%
min}}},\Omega\!\!\upharpoonright\!\! \mathcal{H}_\text{{\scriptsize {min}}}%
)$ explicitly in terms of subspaces of the eigenspaces for $(\mathcal{H}%
,\Omega)$.

This result can be extended to the case in which $\mathcal{H}$ is
infinite-dimensional and $\Omega$ has pure point spectrum, say
$\{ \lambda_\alpha \}_{\alpha = 1}^\infty$ (Theorem \ref{thmSpectrum}). In this
case, the class of polynomial functions $p$ of the operator $\Omega$ of
degree less than $n$, which was sufficient for the finite-dimensional case,
must be expanded to include all continuous functions $f(\Omega)$ as
understood in the classical functional calculus. Applying all continuous
functions of $\Omega$ to all vectors in $H_1$ and taking the closure gives
the orbit ${\mathcal{O}}_{\Omega}(H_1)$, and the projections $%
\pi_\alpha(H_1) $ onto the eigenspaces are contained in this orbit. We
obtain again the spectral decomposition \eqref{HminSpectrum1a}, in which $%
n=\infty$.

A typical frequency operator $\Omega$ for a conservative dynamical system
does not have pure point spectrum, and we now face the problem of extending
this construction of $\mathcal{H}_\text{{\scriptsize {min}}}$ to the case of
general spectrum. The fairly simple construction of $\mathcal{H}_\text{%
{\scriptsize {min}}}$ we have discussed for pure point spectrum provides us
with the correct principle:

\begin{principle}[extension by modes of $\Omega$ via observable variables]
\label{rule1} The minimal conservative extension of $(H_1,\Omega\!\!%
\upharpoonright\!\! H_1,a_1(t))$ is generated by linear superposition,
within the given master system $(\mathcal{H},\Omega)$, by all of the modes
that appear in the eigenmode decomposition, into eigenmodes of $\Omega$, of
any of the states of $H_1$.
\end{principle}

When $\Omega$ has continuous spectrum, its modes are no longer finite-norm (finite-energy) states---they no longer exist as elements of the Hilbert space $\mathcal{H}$. In this case, it one can replace the projections $\pi_\alpha(H_1)$ of $H_1$
to the eigenspaces of $\Omega$ with a set of spectral projections associated
with $\Omega$: 
\begin{equation}  \label{commutingprojections1a}
\left\{ \pi(H_1) : \pi = \int_\Delta \, dE_\lambda \text{ for some interval $%
\Delta$ of $\mathbb{R}$} \right\},
\end{equation}
in which $d E_\lambda$ is the spectral resolution of the identity associated with $\Omega$.  This set generates $\mathcal{H}_\text{{\scriptsize {min}}}$ by linear
combination and closure. Of course, the spaces $\pi(H_1)$ are in general no
longer orthogonal to each other for two different choices of the projection $%
\pi$, so we no longer have an orthogonal decomposition as in equation \eqref{HminSpectrum1a}. The projections can be localized to include only spectral
intervals of length $\epsilon$ for arbitrarily small $\epsilon$, so that one
approaches spectrally localized projections, nearly representing eigenmode
spaces, as $\epsilon\to0$. However, the projections no longer make sense for 
$\epsilon=0$ (unless the spectrum has no continuous part).

Fortunately, one need not abandon the use of modes altogether when dealing
with continuous spectra. Just as the Laplace operator $-\!\sum_i%
\partial_{x_ix_i}$ has the extended states $e^{i\lambda x}$ for its modes,
which generate all sufficiently regular functions through integral
superposition, a proper treatment of modes of $\Omega$ and decomposition of
states into these modes is accomplished by a furnishing of $\mathcal{H}$: $%
\mathcal{H}_+ \subset \mathcal{H }\subset \mathcal{H}_-$. The modes lie in
the larger Hilbert space $\mathcal{H}_-$, endowed with a smaller norm, with
respect to which $\mathcal{H}$ is dense in $\mathcal{H}_-$. All elements of
the smaller space $\mathcal{H}_+$, which is dense in $\mathcal{H}$, are
represented as integral superpositions of the modes: 
\begin{equation}
v = \int \Psi_v(\lambda) d\mu,
\end{equation}
in which $d\mu$ is a spectral measure for $\Omega$, $\Psi$ is a $d\mu$%
-measureable function with values in $\mathcal{H}_-$, and $\Psi(\lambda)$ is
a mode for $\Omega$ for the frequency $\lambda$. This means that $\langle
\Psi(\lambda) | \Omega | v \rangle = \lambda \langle \Psi(\lambda) | v
\rangle$ whenever all of these objects are defined. See, for example, \cite%
{poerschke}.

With this structure, each state $v \in H_1\cap\mathcal{H}_+$ is decomposed
into its modes $\Psi_v(\lambda)$, where $\lambda$ runs over all spectral
values. Integral superpositions of these modes then generate $\mathcal{H}_%
\text{{\scriptsize {min}}}$: 
\begin{equation}  \label{HminSpectrumCont1a}
\mathcal{H}_\text{{\scriptsize {min}}} = \left\{ \int \!\Psi(\lambda) d\mu :
\Psi \in L^2(\mathbb{R},\mathcal{H}_-,d\mu), \, \forall \lambda\in\mathbb{R}%
\, \exists v\in H_1,\, \Psi(\lambda) = \Psi_v(\lambda) \right\}.
\end{equation}
A rigorous treatment of generalized modes is quite technical; we do not
pursue it in this work but leave it for a forthcoming work in which we treat
unbounded coupling.

\medskip

\noindent {\bfseries \ref{subsecReconstructiblilty2}.2 Generating the
conservative extension from the coupling channels.} \ The role of the coupling operator $\Gamma$ and the space of hidden variables $\mathcal{H}\ominus H_1$ is not emphasized in the construction of $\mathcal{H}_\text{{\scriptsize {min}}}$ from $H_1$ by the action of $\Omega$. We shall now show an alternative way to generate $\mathcal{H}_\text{{\scriptsize {min}}}$, provided $H_1 = H_{1c}$, that is, provided that $H_1$ has no $\Omega$-invariant subspaces. This is by the action of $\Omega$ on $\mathrm{Ran} \,\Gamma$ and $\mathrm{Ran}\,\Gamma^\dag$, or by the action of $\Omega$ on the coupling channels (see definition \ref{defCouplingChannel}).

We shall see that the minimal reconstructible system described in Theorem %
\ref{thmReconstructibility} is obtained by linear superposition of the
projections of the ranges of $\Gamma$ and $\Gamma^\dag$ onto the eigenspaces
of $\Omega$. This can also be expressed as linear superposition of those
modes of $\Omega$ that are present in the mode decompositions of all vectors
in $\mathrm{Ran}\,\Gamma$ and $\mathrm{Ran}\,\Gamma^\dag$.

To understand why this is true, let us decompose $\Omega$ into a diagonal
part representing the internal dynamics of the observable and hidden
variables and a part representing the coupling: 
\begin{equation}
\Omega = \mathring\Omega + \mathring\Gamma, \quad \mathring\Omega = \left[ 
\begin{array}{cc}
\Omega_1 & 0 \\ 
0 & \Omega_2%
\end{array}
\right], \quad \mathring\Gamma = \left[ 
\begin{array}{cc}
0 & \Gamma \\ 
\Gamma^\dagger & 0%
\end{array}
\right].
\end{equation}
Since $\mathrm{Ran}\,\mathring\Gamma$ is contained in ${%
\mathcal{O}}_{\Omega}(\mathrm{Ran}\,\mathring\Gamma)$ and $\Omega =
\mathring\Omega + \mathring\Gamma$, the operator $\Omega$ generating the
orbit can be replaced by $\mathring\Omega$:  Recalling the definition of the orbit of a subset of a Hilbert space under the action of two operators \eqref{dorbit2}, we obtain
\begin{equation}
{\mathcal{O}}_{\Omega }(\mathrm{Ran}\,\mathring{\Gamma}) = {\mathcal{O}}%
_{\Omega ,\mathring{\Gamma}}(\mathrm{Ran}\,\mathring{\Gamma}) = {\mathcal{O}}%
_{\mathring{\Omega},\mathring{\Gamma}} (\mathrm{Ran}\,\mathring{\Gamma}) = {%
\mathcal{O}}_{ \mathring{\Omega}}(\mathrm{Ran}\,\mathring{\Gamma}).
\end{equation}
Now, by the decoupling of the action of $\mathring\Omega$ with respect to
the decomposition $H_1\oplus H_2$ and the splitting $\mathrm{Ran}%
\,\mathring\Gamma = \mathrm{Ran}\,\Gamma \oplus \mathrm{Ran}\,\Gamma^\dag$,
we obtain a simple characterization of this orbit: 
\begin{equation}
{\mathcal{O}}_{\mathring{\Omega}}(\mathrm{Ran}\,\mathring{\Gamma}) = {%
\mathcal{O}}_{\mathring{\Omega}}(\mathrm{Ran} \,\Gamma )\oplus {\mathcal{O}}%
_{\mathring{\Omega}}(\mathrm{Ran}\,\Gamma ^{\dagger }) = {\mathcal{O}}%
_{\Omega_1}(\mathrm{Ran}\,\Gamma) \oplus {\mathcal{O}}_{\Omega_2}(\mathrm{Ran%
}\,\Gamma^\dag).
\end{equation}
Assuming that $H_1 = H_{1c}$, or, equivalently, that $H_1$ has no nontrivial 
$\Omega$-invariant subspace (see Theorem \ref{thmReconstructibility}, part (%
\textit{i})), we obtain 
\begin{equation}  \label{OrbitGamma}
\mathcal{H}_\text{{\scriptsize {min}}} = {\mathcal{O}}_{\Omega}(\mathrm{Ran}%
\,\mathring\Gamma).
\end{equation}

We may now adapt our previous discussion concerning the construction of $%
\mathcal{H}_\text{{\scriptsize {min}}}$ from $H_1$ in order to understand
the construction of $\mathcal{H}_\text{{\scriptsize {min}}}$ from $\mathrm{%
Ran}\,\mathring\Gamma$ within the spectral structure of $(\mathcal{H}%
,\Omega)$ simply be replacing $H_1$ in the arguments with $\mathrm{Ran}%
\,\mathring\Gamma$.

\begin{principle}[extension by modes of $\Omega$ via coupling channels]
\label{rule2} If $H_1$ contains no $\Omega$-invariant subspace, then the
minimal conservative extension of $(H_1,\Omega\!\!\upharpoonright\!\!
H_1,a_1(t))$ is generated by linear superposition, within the given master
system $(\mathcal{H},\Omega)$, by all of the modes that appear in the
eigenmode decomposition, into eigenmodes of $\Omega$, of any of the states
in the ranges of $\Gamma$ and $\Gamma^\dagger$.
\end{principle}

The representations \eqref{HminSpectrum1a} and \eqref{HminSpectrumCont1a}
with $\mathrm{Ran}\,\mathring\Gamma$ replacing $H_1$ are valid for the pure
point and general cases, repectively.
In the case of pure point spectrum, in particular for finite systems, we can
reformulate Rule \ref{rule2} in terms of a characterization of
reconstructiblity: 

\vspace{1ex}

\noindent
\hspace*{0.05\textwidth}
\parbox{0.9\textwidth}{\em
$(\mathcal{H}\!=\! H_1\oplus H_2,\Omega)$ is
reconstructible if and only if, on each mode of $\Omega$, $\mathring\Gamma$
does not vanish. }

\vspace{1ex}

\medskip

\noindent {\bfseries \ref{subsecReconstructiblilty2}.4 Generating $H_1$ and $%
H_2$ from $\Gamma$.} \ These rules can be applied as well to the two
components of the reconstructible system $(\mathcal{H}^{\prime }\!=\!
H_{1c}\oplus H_{2c},\ \Omega\!\!\upharpoonright\!\!\mathcal{H}^{\prime })$
from the operators of their internal dynamics, $\Omega_1$ and $\Omega_2$. Recall
from part (\textit{i}) of Theorem \ref{thmReconstructibility} that the
``coupled" part $H_{1c}$ of $H_1$ is generated through the action of the
frequency operator $\Omega_1$ on the range of $\Gamma$. We obtain therefore,
by the same reasoning as before, results on the construction of $H_{1c}$ by
superposition of the modes of $\Omega_1$ obtained from the projections of $%
\mathrm{Ran}\,\Gamma$ onto the eigenspaces of $\Omega_1$. Of course, this
applies equally to the construction of $H_{2c}$ by modes of $\Omega_2$.

\begin{principle}[Generating $H_{1c}$ and $H_{2c}$]
\label{rule3} The ``coupled" part $H_{ic}$ of $H_{i}$ is generated by linear
superposition, within the system $(H_i,\Omega_i)$, by all of the modes that
appear in the eigenmode decomposition, into modes of $\Omega_i$, of any of
the states in $\mathrm{Ran}\,\Gamma$ ($i=1$) or $\mathrm{Ran}\,\Gamma^\dag$ (%
$i=2$).
\end{principle}

In the case of pure point spectrum, 
\begin{equation}
H_{1c} = {\mathcal{O}}_{\Omega_1} (\mathrm{Ran}\,\Gamma) =
\bigoplus\limits_{\alpha=1}^n \pi_{1\alpha}(\mathrm{Ran}\,\Gamma),
\end{equation}
\begin{equation}
H_{2c} = {\mathcal{O}}_{\Omega_2} (\mathrm{Ran}\,\Gamma^\dag) =
\bigoplus\limits_{\alpha=1}^m \pi_{2\alpha}(\mathrm{Ran}\,\Gamma^\dag),
\end{equation}
in which $\pi_{1\alpha}$ and $\pi_{2\alpha}$ are projections onto the
eigenspaces of $\Omega_1$ in $H_1$ and $\Omega_2$ in $H_2$, respectively. If 
$\Omega_i$ has continuous spectrum, then, as in our previous discussion, we
may replace projections onto the eigenspaces by general spectral
projections, that is, those projections that commute with $\Omega_i$ (see %
\eqref{commutingprojections1a}). The discussion of generalized modes and a
construction of the form \eqref{HminSpectrumCont1a} for $H_{ic}$ is also
applicable.

\medskip

\noindent {\bfseries \ref{subsecReconstructiblilty2}.5 Summary and theorem.}
\ The results of the discussion are collected in the following theorem and
proved in Section \ref{sectionProofs} (see the proof of Theorem \ref%
{thmSpectrum2}).

We have made the statement about decomposition into modes rigorous in part (%
\textit{i}) of the theorem for the case of pure point spectrum, in which the
modes are genuine elements of $\mathcal{H}$; a weaker rigorous statement in
which the modes are replaced by arbitrary spectral projections, is given for
general spectrum in part (\textit{ii}).

Theorem \ref{thmMultiplicity} is one of our main results; it summarizes our conclusions about bounding the multiplicity of the frequency operators $\Omega\!\!\upharpoonright\!\!%
\mathcal{H}_\text{{\scriptsize {min}}}$ and $\Omega_{ic} =
\Omega_i\!\!\upharpoonright\!\! H_{ic}$ by the rank of $\Gamma$. In
particular, the result for $\Omega_{2c}$ states that the number of strings
needed to construct the minimal extension is bounded by the number of
coupling channels.

\begin{theorem}[spectral representation of the minimal extension]
\label{thmSpectrum} Let a conservative system $(\mathcal{H},\Omega )$ be
given, and let $(\mathcal{H}_\text{{\scriptsize {min}}}\hspace{-0.3em}%
\subset \hspace{-0.3em} \mathcal{H},\Omega \!\!\upharpoonright \!\!%
\mathcal{H}_\text{{\scriptsize {min}}})$ be the minimal conservative
extension of the open system $(H_1,\Omega\!\!\upharpoonright\!\!
H_1,a_1(t))$ obtained by projecting the dynamics of $(\mathcal{H},\Omega )$
onto the subspace $H_{1}\subset \mathcal{H}$. The coupling operator $\Gamma$
is assumed to be bounded.  (By the ``projection" of a subset of a Hilbert space onto a subspace, we refer to the image of the orthogonal projection operator in the Hilbert space onto the subspace.)

\begin{enumerate}
\item If $\Omega$ has pure point spectrum, then $\mathcal{H}_\text{%
{\scriptsize {min}}}$ is the closure of the linear span of the projections
of $H_1$ onto the eigenspaces of $\Omega$ in $\mathcal{H}$: 
\begin{equation}  \label{HminSpectrum1b}
\mathcal{H}_\text{{\scriptsize {min}}} = {\mathcal{O}}_{\Omega}(H_1) =
\bigoplus\limits_{\alpha=1}^n \pi_\alpha(H_1).
\end{equation}
If, in addition $H_1$ contains no nontrivial $\Omega$-invariant subspace,
then $\mathcal{H}_\text{{\scriptsize {min}}}$ is the closure of the linear
span of the projections of $\mathrm{Ran}\,\mathring\Gamma$ onto the
eigenspaces of $\Omega$: 
\begin{equation}  \label{HminSpectrum2}
\mathcal{H}_\text{{\scriptsize {min}}} = {\mathcal{O}}_{\Omega}(\mathrm{Ran}%
\,\mathring\Gamma) = \bigoplus\limits_{\alpha=1}^n \pi_\alpha(\mathrm{Ran}%
\,\mathring\Gamma).
\end{equation}
Here, $n$ may be equal to infinity.

\item $\mathcal{H}_\text{{\scriptsize {min}}}$ is the closure of the linear
span of the projections 
\begin{equation}  \label{commutingprojections1b}
\left\{ \pi(H_1) : \pi = \int_\Delta \, dE_\lambda \text{ for some interval $%
\Delta$ of $\mathbb{R}$} \right\}.
\end{equation}
If $H_1$ contains no nontrivial $\Omega$-invariant subspace, then $\mathcal{H%
}_\text{{\scriptsize {min}}}$ is the closure of the linear span of the
projections 
\begin{equation}  \label{commutingprojections2}
\left\{ \pi(\mathrm{Ran}\,\mathring\Gamma) : \pi = \int_\Delta \, dE_\lambda 
\text{ for some interval $\Delta$ of $\mathbb{R}$} \right\}.
\end{equation}
For arbitrary $\epsilon>0$, the set of projections can be restricted to
those that vanish outside some spectral interval $\Delta$ of length $%
\epsilon $.

\item Let $P_1$ denote orthogonal projection onto $H_1$ within $\mathcal{H}$%
. If $\Omega$ has pure point spectrum, then $\mathcal{H}_\text{{\scriptsize {%
min}}}=\mathcal{H}$ if and only if $P_1(\phi) \not= 0$ for each eigenmode $%
\phi$ of $\Omega$. %
If, in addition, $H_1$ has no nontrivial $\Omega$-invariant subspace, then $%
\mathcal{H}_\text{{\scriptsize {min}}}=\mathcal{H}$ if and only if $%
\mathring\Gamma(\phi) \not= 0$ for each eigenmode $\phi$ of $\Omega$. %
%
In fact, $(\mathcal{H}\!=\! H_1\oplus H_2,\Omega)$ is reconstructible if and
only if $\mathring\Gamma(\phi) \not= 0$ for each eigenmode $\phi$ of $\Omega$%
.
\end{enumerate}
\end{theorem}

\begin{theorem}[spectral representation of the hidden variables]
\label{thmSpectrum2} Let the hypotheses of Theorem \ref{thmSpectrum}
continue to hold.

\begin{enumerate}
\item If $\Omega_2$ has pure point spectrum, then $H_{2c}$ is the closure of
the linear span of the projections of $\mathrm{Ran}\,\Gamma^\dag$ onto the
eigenspaces of $\Omega_2$ in $H_2$: 
\begin{equation}  \label{H2Spectrum}
H_{2c} = {\mathcal{O}}_{\Omega_2}(\mathrm{Ran}\,\Gamma^\dag) =
\bigoplus\limits_{\alpha=1}^n \pi_{2\alpha}(\mathrm{Ran}\,\Gamma^\dag).
\end{equation}
Here, $n$ may be equal to infinity.

\item $H_{2c}$ is the closure of the linear span of the projections 
\begin{equation}
\left\{ \pi(\mathrm{Ran}\,\Gamma^\dag) : \pi = \int_\Delta \, dE_{2,\lambda} 
\text{ for some interval $\Delta$ of $\mathbb{R}$} \right\}.
\end{equation}
For arbitrary $\epsilon>0$, the set of projections can be restricted to
those that vanish outside some spectral interval $\Delta$ of length $%
\epsilon $.

\item If $\Omega_2$ has pure point spectrum, then $H_{2c} = H_2$ if and only
if $\Gamma^\dag(\phi) \not= 0$ for each eigenmode $\phi$ of $\Omega_2$.
\end{enumerate}
\end{theorem}

\noindent
The following theorem is a corollary to the preceding theorems.  It is one of our main results, which we have alluded to in the introduction and in the construction of the toy model of a solid with frozen degrees of freedom in Section \ref{sectionFrozen}.  The second part shows that the rank of the coupling operator bounds the number of extending strings needed in the minimal conservative extension of an open system.

\begin{theorem}[bound on number of strings]
\label{thmMultiplicity} \hspace{1pt}

\begin{enumerate}
\item The multiplicity of each spectral value $\lambda$ of $%
\Omega\!\!\upharpoonright\!\!\mathcal{H}_\text{{\scriptsize {min}}}$ is
bounded above by the dimension of $H_1$ and, if $H_{1c}=H_1$, by twice the
rank of $\Gamma$: 
\begin{equation}
\mathrm{multiplicity}\,(\lambda) \leq \min\{\dim(H_1),2\,\mathrm{rank}%
\,(\Gamma)\}.
\end{equation}

\item The multiplicity of each spectral value $\lambda$ of $%
\Omega_i\!\!\upharpoonright\!\! H_{ic}$ is bounded above by the rank of $%
\Gamma$: 
\begin{equation}
\mathrm{multiplicity}\,(\lambda) \leq \mathrm{rank}\,(\Gamma).
\end{equation}
For $i=2$, this states that the number of abstract strings needed to extend $%
(H_1,\Omega\!\!\upharpoonright\!\! H_1,a_1(t))$ minimally to a
conservative system is no greater than the number of coupling channels
between $H_1$ and $H_2$.
\end{enumerate}
\end{theorem}


\section{Decomposition of coupled systems}

\label{sectionDecomposition}

It can happen that a given open system splits into two or more smaller
independent subsystems, that are decoupled from each other, leading to a natural
simplifying decomposition. But, depending on the choice of coordinates, such
a natural decomposition may not be evident right away. We ask then if there
is a systematic way to find such a decomposition. In this section we intend to
answer this question, at least under tractable conditions; the most general conditions are treated in Section \ref{sectionProofs}.

Let us return to our observable open system: 
\begin{equation}
\partial _{t}v_{1}(t)=-\mathrm{i}\Omega _{1}v_{1}(t)-\int_{0}^{\infty
}a_{1}(\tau )v_{1}(t-\tau )\,\,\mathrm{d}\tau +f_{1}(t)\quad \text{in }H_{1},
\label{reduced1again}
\end{equation}%
and suppose that there is a subspace of observable variables $H_{1}^{\prime
}\subset H_{1}$ such that an observer confined to this subspace experiences
no influence from the rest of the observable space, that is, $H_{1}^{\prime }$ is
decoupled from $H_{1}^{\prime \prime }=H_{1}\ominus H_{1}^{\prime }$ under
the dynamics of \eqref{reduced1again}. More precisely, let $\pi _{1}^{\prime
}$ be the orthogonal projection onto $H_{1}^{\prime }$ in $H_{1}$ and $\pi
_{1}^{\prime \prime }=\mathbb{I}_{H_{1}}\!-\!\pi _{1}^{\prime }$ the
projection onto $H_{1}^{\prime \prime }$, and let $v_{1}^{\prime }(t)=\pi
_{1}^{\prime }v_{1}(t)$ and $f_{1}^{\prime }(t)=\pi _{1}^{\prime }f_{1}(t)$.
Then the decoupling of $H_{1}^{\prime }$ means that 
\begin{equation}
\pi _{1}^{\prime }\Omega _{1}\pi _{1}^{\prime \prime }=0\qquad \text{and}%
\qquad \pi _{1}^{\prime }a_{1}(t)\pi _{1}^{\prime \prime }=0\quad \text{for
all $t$},
\end{equation}%
so that $v^{\prime }(t)$ satisfies a dynamical equation within $%
H_{1}^{\prime }$, with no input from $H_{1}^{\prime \prime }$:%
\begin{equation}
\partial _{t}v_{1}^{\prime }(t)=-\mathrm{i}\Omega _{1}v_{1}^{\prime
}(t)-\int_{0}^{\infty }a_{1}(\tau )v_{1}^{\prime }(t-\tau )\,\,\mathrm{d}%
\tau +f_{1}^{\prime }(t)\quad \text{in }H_{1}^{\prime }.
\label{reduced1prime}
\end{equation}%
The question then arises: Does this imply the reciprocal condition that that 
$H_{1}^{\prime \prime }$ evolves independently of $H_{1}^{\prime }$? In
other words, if $H_{1}^{\prime }$ is not influenced by $H_{1}^{\prime \prime
}$, then does it follow that $H_{1}^{\prime \prime }$ is not influenced by $%
H_{1}^{\prime }$? We shall prove that the answer is affirmative. This means
that $\pi _{1}^{\prime \prime }a_{1}(t)\pi _{1}^{\prime }=0$ also, so that
the splitting $H_{1}=H_{1}^{\prime }\oplus H_{1}^{\prime \prime }$ is
preserved by $a_{1}(t)$, for all $t$. In this case, we have a decoupling of
the open system \eqref{reduced1again} into two independent open systems, so
that%
\begin{equation}
\partial _{t}v_{1}^{\prime \prime }(t)=-\mathrm{i}\Omega _{1}v_{1}^{\prime
\prime }(t)-\int_{0}^{\infty }a_{1}(\tau )v_{1}^{\prime \prime }(t-\tau )\,\,%
\mathrm{d}\tau +f_{1}^{\prime \prime }(t)\quad \text{in }H_{1}^{\prime
\prime }  \label{reduced1doubleprime}
\end{equation}%
also holds.

Such decoupling of open systems is easy to understand \emph{from the point
of view of the minimal conservative extension} $(\mathcal{H},\Omega )$ of $%
(H_{1},\Omega _{1},a_{1}(t))$. If $(\mathcal{H}^{\prime },\Omega ^{\prime })$
and $(\mathcal{H}^{\prime \prime },\Omega ^{\prime \prime })$ are the
minimal conservative extensions of the open systems in $H_{1}^{\prime }$ and 
$H_{1}^{\prime \prime }$, then $(\mathcal{H}:=\mathcal{H}^{\prime }\oplus 
\mathcal{H}^{\prime \prime },\Omega :=\Omega ^{\prime }\oplus \Omega
^{\prime \prime })$ is the minimal conservative extension of $(H_{1},\Omega
_{1},a_{1}(t))$. \emph{The decoupling of the open system (\ref{reduced1again}%
) is tantamount to the existence of a projection in its conservative
extension $\mathcal{H}$, namely the projection $\pi ^{\prime }$ onto $%
\mathcal{H}^{\prime }$, that commutes both with $\Omega $ as well as with
the projection $P_{1}$ onto $H_{1}$.} This is the content of Theorem \ref%
{thmOpen} (and Theorem \ref{thmOpen2} in Section \ref{sectionProofs}) below.

The structure of this decomposition and its implications for the
decomposition of $\Omega _{1}$, $\Omega _{2}$, and $\Gamma $ can be seen in
a four-fold decomposition of $(\mathcal{H},\Omega )$.  Denote by $%
P_{2}=\mathbb{I}-P_{1}$ the projection onto $H_{2}$ and by $\pi ^{\prime
\prime }=\mathbb{I}-\pi ^{\prime }$ the projection onto $\mathcal{H}^{\prime
}$. We note that $P_{1}$ and $\pi ^{\prime }$ commute if and only if $%
\mathcal{H}$ admits the (orthogonal) decomposition 
\begin{equation}
\mathcal{H}=H_{1}^{\prime }\oplus H_{1}^{\prime \prime }\oplus H_{2}^{\prime
}\oplus H_{2}^{\prime \prime },  \label{decomposition}
\end{equation}%
where the components are, respectively, the images of the projections $\pi
^{\prime }P_{1}$, $\pi ^{\prime \prime }P_{1}$, $\pi ^{\prime }P_{2}$, and $%
\pi ^{\prime \prime }P_{2}$. With respect to the decomposition (\ref%
{decomposition}), the operator $\Omega $ has the form 
\begin{equation}
\Omega =\left[ 
\begin{array}{cccc}
\Omega _{1}^{\prime } & 0 & \Gamma ^{\prime } & 0 \\ 
0 & \Omega _{1}^{\prime \prime } & 0 & \Gamma ^{\prime \prime } \\ 
\Gamma ^{\prime \dagger } & 0 & \Omega _{2}^{\prime } & 0 \\ 
0 & \Gamma ^{\prime \prime \dagger } & 0 & \Omega _{2}^{\prime \prime }%
\end{array}%
\right] ,  \label{blockform}
\end{equation}%
in which the splittings $H_{1}=H_{1}^{\prime }\oplus H_{1}^{\prime \prime }$
and $H_{2}=H_{2}^{\prime }\oplus H_{2}^{\prime \prime }$ \emph{%
simultaneously diagonalize} $\Omega _{1}$, $\Omega _{2}$, and $\Gamma $.

This type of system decoupling of a conservative extension $(\mathcal{H}%
,\Omega)$ of a given open system $(H_1,\Omega_1,a_1(t))$ we call \emph{%
s-invariant} (for system-invariant) with respect to $H_1$.

\begin{definition}[s-invariant decomposition]
Let a conservative system $(\mathcal{H},\Omega )$ be given, along with an
open subsystem obtained by projecting the dynamics onto a subspace $%
H_{1}\subset \mathcal{H}$, and let $P_{1}$, $\Omega _{1}$, $\Omega _{2}$,
and $\Gamma $ be defined as before. A decomposition $\mathcal{H=H^{\prime
}\oplus H^{\prime \prime }}$, with projection $\pi ^{\prime }$ onto $%
\mathcal{H}^{\prime }$, is called \emph{s-invariant with respect to $H_{1}$
(or $P_{1}$)} if the following equivalent conditions hold:

\begin{enumerate}
\item $\pi^{\prime }$ commutes with $\Omega$ and $P_1$;

\item $\mathcal{H}^{\prime }$ (or, equivalently, $\mathcal{H}^{\prime \prime
}$) is of the form $\mathcal{H}^{\prime }= H^{\prime }_1\oplus H^{\prime }_2$, where $H^{\prime }_1\subset H_1$ and $H^{\prime }_2\subset H_2$, and $\mathcal H'$ is
invariant under $\Omega$ ($\orbit{\Omega}(\mathcal H') = \mathcal H'$).
\end{enumerate}

An \emph{s-invariant decomposition of a conservative extension of $%
(H_1,\Omega_1,a_1(t))$} is understood to be s-invariant with respect to the
subspace $H_1$.
\end{definition}

\begin{theorem}[decoupling criterion]
\label{thmOpen} Let the open system (\ref{reduced1again}) characterized by
the triple $(H_{1},\Omega _{1},a_{1}(t))$ be given, along with a subspace $%
H_{1}^{\prime }\subset H_{1}$. The following are equivalent:

\begin{enumerate}
\item The projected dynamics of the open system onto $H^{\prime }_1$ is
not influenced by the dynamics of $H_1\ominus H_1'$, that is, the dynamical equation \eqref{reduced1prime} holds.

\item The minimal conservative extension $(\mathcal{H},\Omega)$ of the open
system admits an s-invariant splitting $\mathcal{H = H^{\prime }\oplus
H^{\prime \prime }}$ such that $H^{\prime }_1 = H_1\cap\mathcal{H}^{\prime }$, that is, the block-diagonal form \eqref{blockform} for $\Omega$ holds.
\end{enumerate}
\end{theorem}

Notice that part ({\itshape ii}) implies that $a_1(t)$ is diagonal with
respect to the decomposition $H_1 = H^{\prime }_1 \oplus H^{\prime }_2$ so
that both \eqref{reduced1prime} and \eqref{reduced1doubleprime} hold.
Therefore, by the theorem, \eqref{reduced1prime} (or %
\eqref{reduced1doubleprime}) is equivalent to (\ref{reduced1prime},\ref%
{reduced1doubleprime}).

Theorem \ref{thmDecomposition} involves
the relation between s-invariant decompositions of conservative systems and
the singular values of the coupling operator. We begin with a treatment of an
{\em arbitrary} countable orthogonal decomposition of $H_{1}$ and $H_{2}$ that
is invariant under the internal actions given by $\Omega _{1}$ and
$\Omega_{2}$ but does not necessarily correspond to an s-invariant decomposition: 
\begin{equation}
H_{i}=\textstyle{\bigoplus\limits_{\alpha =1}^{n_{i}}}H_{i\alpha },\quad 
\mathbb{I}_{H_{i}}=\sum\limits_{\alpha =1}^{n_{i}}\pi _{i\alpha },\qquad
i=1,2,  \label{Hdecomposition}
\end{equation}%
where $\pi _{i\alpha }$, $i=1,2$, are the orthogonal projections onto the
the subspaces $H_{i\alpha }$ and the $n_{i}$ are allowed to be $\infty $.

The frequency and coupling operators split as follows: 
\begin{equation}
\Omega _{i}=\sum_{\alpha =1}^{n_{i}}\Omega _{i\alpha },
\end{equation}
\begin{equation}
\Gamma \,=\,\mathbb{I}_{H_{1}}\Gamma \,\mathbb{I}_{H_{2}}\,
=\,\sum_{\alpha =1}^{n_{1}}\sum_{\beta =1}^{n_{2}}\Gamma _{\alpha \beta }\,,
\quad
\Gamma _{\alpha \beta }=\pi _{1\alpha }\Gamma \pi _{2\beta }.
\end{equation}
Given $\Gamma _{\alpha \beta }=\pi _{1\alpha }\Gamma \pi _{2\beta }$, we
have also $\Gamma _{\alpha \beta }^{\dagger }=\pi _{2\beta }\Gamma ^{\dagger
}\pi _{1\alpha }$. In block-matrix form, with $n_{1}=2$ and $n_{2}=3$, $%
\Omega $ has the form 
\begin{equation}
\Omega =\left[ 
\begin{array}{cc}
\Omega _{1} & \Gamma \\ 
\Gamma ^{\dagger } & \Omega _{2}%
\end{array}%
\right] =\left[ 
\begin{array}{ccccc}
\Omega _{11} & 0 & \Gamma _{11} & \Gamma _{12} & \Gamma _{13} \\ 
0 & \Omega _{12} & \Gamma _{21} & \Gamma _{22} & \Gamma _{23} \\ 
\Gamma _{11}^{\dag } & \Gamma _{21}^{\dag } & \Omega _{21} & 0 & 0 \\ 
\Gamma _{12}^{\dag } & \Gamma _{22}^{\dag } & 0 & \Omega _{22} & 0 \\ 
\Gamma _{13}^{\dag } & \Gamma _{23}^{\dag } & 0 & 0 & \Omega _{23}%
\end{array}%
\right] .
\end{equation}%
We refine our decomposition of $\Gamma $ through its singular-value
decomposition, aided by the Hilbert-space isomorphism between the range of $%
\Gamma $ in $H_{1}$ and the range of $\Gamma ^{\dagger }$ in $H_{2}$,
described in Section \ref{secExtension} (page \pageref{U}): 
\begin{equation}
U:\mathrm{Ran}\,\Gamma \rightarrow \mathrm{Ran}\,\Gamma ^{\dagger }.
\end{equation}

For the sake of technical simplicity we restrict discussion in this work to
the situation in which the spectrum of $\Gamma _{R}\Gamma _{R}^{\dagger }$
consists only of eigenvalues and their accumulation points, and state the
result, Theorem \ref{thmDecomposition}, for this case. Its proof as well as
a partial generalization of it is given in Section \ref{sectionProofs}
(Theorem \ref{thmDecomposition2}). The eigenspaces of $\Gamma _{R}\Gamma
_{R}^{\dagger }$ and $\Gamma _{R}^{\dagger }\Gamma _{R}$ for the same
eigenvalue are identified isometrically through $U$. Let $r=\mathrm{rank}%
\,\Gamma =\mathrm{rank}\,\Gamma ^{\dagger }$, which is allowed to be $\infty 
$, and let $\{g_{q}\}_{q=1}^{r}$ be an orthonormal basis for $\mathrm{Ran}%
\,\Gamma $ consisting of eigenvectors of $\Gamma _{R}\Gamma _{R}^{\dagger }$
with eigenvalues $\gamma _{q}>0$, and put $g_{q}^{\prime }=Ug_{q}$. Each of
the vectors $g_{q}$ and $g_{q}^{\prime }$ is an eigenvector (or generalized
eigenvector if $\Gamma $ is unbounded) of respectively $\Gamma _{R}\Gamma
_{R}^{\dag }$ and $\Gamma _{R}^{\dag }\Gamma _{R}$.

With the help of the Dirac notation, in which $\left\vert g_{q}\right\rangle 
$ indicates the vector $g_{q}$ and $\left\langle g_{q}\right\vert $ the
linear functional of projection onto $g_{q}$, we can write $\Gamma \Gamma
^{\dagger }$ and its adjoint as a sum of rank-one operators: 
\begin{equation}
\Gamma \Gamma ^{\dag }=\sum_{q=1}^{r}\gamma _{q}\left\vert
g_{q}\right\rangle \left\langle g_{q}\right\vert ,\qquad \Gamma ^{\dag
}\Gamma =\sum_{q=1}^{r}\gamma _{q}\left\vert g_{q}^{\prime }\right\rangle
\left\langle g_{q}^{\prime }\right\vert .  \label{GammaGamma}
\end{equation}%
(If $\left\vert g_{q}\right\rangle $ is a genuine eigenvector of $\Gamma
_{R}\Gamma _{R}^{\dag }$ as we assumed, then $\left\vert g_{q}\right\rangle
\in H_{1}$; if it is generalized, then $\left\vert g_{q}\right\rangle \in %
\left[ H_{1}\right] _{-}$, where $\left[ H_{1}\right] _{+}\subset
H_{1}\subset \left[ H_{1}\right] _{-}$ is a proper furnishing of $H_{1}$).
If all $\gamma _{q}$ are different, then the representations (\ref%
{GammaGamma}) are unique. If some of the $\gamma _{q}$ coincide, then they
are not, and we may choose orthonormal eigenvectors arbitrarily from the
eigenspace. With this structure, $\Gamma _{R}$ can be written as a sum of
linearly independent rank-one operators: 
\begin{equation}
\Gamma _{R}=\sum_{q=1}^{r}\Gamma _{q},\qquad \Gamma _{q}=\sqrt{\gamma _{q}}%
\left\vert g_{q}\right\rangle \left\langle g_{q}^{\prime }\right\vert ,\quad
\gamma _{q}>0,\   \label{Gamma}
\end{equation}%
\begin{equation}
\left\langle g_{p}\right\vert \left\vert g_{q}\right\rangle =\left\langle
g_{p}^{\prime }\right\vert \left\vert g_{q}^{\prime }\right\rangle =\delta
_{pq}.
\end{equation}%
For each $q=1,\dots ,r$, decompose $|g_{q}\rangle $ and $|g_{q}^{\prime
}\rangle $ with respect to the Hilbert space decompositions %
\eqref{Hdecomposition}: 
\begin{equation}
|g_{q}\rangle =\sum_{\alpha =1}^{n_{1}}|g_{q}^{\alpha }\rangle ,\qquad
|g_{q}^{\prime }\rangle =\sum_{\alpha =1}^{n_{2}}|{g^{\prime }}_{q}^{\alpha
}\rangle .
\end{equation}%
It follows that 
\begin{equation}
\Gamma _{q}\,=\,\sqrt{\gamma _{q}}\sum_{\alpha =1}^{n_{1}}\sum_{\beta
=1}^{n_{2}}|g_{q}^{\alpha }\rangle \big\langle{g^{\prime }}_{q}^{\beta }\big|%
,
\end{equation}%
so that $\Gamma $ is decomposed as 
\begin{equation}
\Gamma =\sum_{\alpha =1}^{n_{1}}\sum_{\beta =1}^{n_{2}}\Gamma _{\alpha \beta
},\qquad \Gamma _{\alpha \beta }=\sum_{q=1}^{r}\sqrt{\gamma _{q}}%
\,\left\vert g_{q}^{\alpha }\right\rangle \big\langle{g^{\prime }}%
_{q}^{\beta }\big\vert.
\end{equation}%
This decomposition shows explicitly the coupling between the components of $%
H_{1}$ and the components of $H_{2}$ in terms of the spectral structure of
the coupling operator. $H_{1\alpha }$ is directly coupled with $H_{2\beta }$
if and only if $\Gamma _{\alpha \beta }=0$. We note, however, that, for a
fixed pair $(\alpha ,\beta )$, the rank-one operators $|g_{q}^{\alpha
}\rangle \langle {g^{\prime }}_{q}^{\beta }|$, for $q=1,\dots ,r$, are not
in general independent, so that $\Gamma _{\alpha \beta }$ may be zero even
if, for some $q$, $g_{q}^{\alpha }$ and ${g^{\prime }}_{q}^{\beta }$ are
both nonzero; in fact, the cardinality of $\{q:g_{q}^{\alpha }\rangle
\langle {g^{\prime }}_{q}^{\beta }|\not=0\}$ may exceed the rank of $\Gamma
_{\alpha \beta }$.

We organize the coupling information by introducing the $n_{1}\times n_{2}$
coupling matrix $M_{\Gamma }$ with entries 
\begin{equation}
\left[ M_{\Gamma }\right] _{\alpha \beta }=\mathrm{rank}\,\Gamma _{\alpha
\beta }.
\end{equation}%
The $\alpha \beta $-component of the coupling matrix can be thought of as
the number of coupling channels between the components $H_{1\alpha }$ and $%
H_{2\beta }$. Rows of $M_{\Gamma }$ containing all zeros indicate components
of $H_{1}$ that split from the rest of the system $H_{1}\oplus H_{2}$, and
columns of all zeros indicate components of $H_{2}$ that split from the rest
of the system. If the subspaces $H_{i\alpha }$ can be reordered in such a
way that $M_{\Gamma }$ attains a diagonal block form (with not necessarily
square blocks), then we see that the system splits into completely decoupled
subsystems, each with a nontrivial component in each of $H_{1}$ and $H_{2}$
made up of components $H_{i\alpha }$ ($i=1,2$). \emph{This leads to an
s-invariant decomposition.} In this case, it is possible to choose the $%
g_{q} $ differently if necessary so that, for $\alpha \beta $ off of the
diagonal blocks, we have $|g_{q}^{\alpha }\rangle \langle {g^{\prime }}%
_{q}^{\beta }|=0$ for all $q$, as we will see.

We now examine s-invariant decompositions in more detail, that is, how $%
\mathcal{H}=H_{1}\oplus H_{2}$ can be decomposed into independently evolving
components of the form $H_{1\alpha }\oplus H_{2\alpha }$: 
\begin{equation}
\mathcal{H}=\textstyle{\bigoplus\limits_{\alpha =1}^{n}}\left( H_{1\alpha
}\oplus H_{2\alpha }\right) ,  \label{CalHdecomposition}
\end{equation}%
where $H_{1\alpha }\oplus H_{2\alpha }$ is invariant under $\Omega $ for
each $\alpha $. This means that $H_{i\alpha }$ is invariant under $\Omega
_{i}$ for $i=1,2$ and $\alpha =1,\dots ,n$ and that the coupling operators $%
\Gamma _{\alpha \beta }$ are equal to zero for $\alpha \not=\beta $. In
other words, \emph{this decomposition simultaneously block-diagonalizes $%
\Omega _{1}$, $\Omega _{2}$, and $\Gamma $}. For $n=2$, for example, $\Omega 
$ has the form 
\begin{equation}
\Omega =\left[ 
\begin{array}{cccc}
\Omega _{11} & 0 & \Gamma _{11} & 0 \\ 
0 & \Omega _{12} & 0 & \Gamma _{22} \\ 
\Gamma _{11}^{\dagger } & 0 & \Omega _{21} & 0 \\ 
0 & \Gamma _{22}^{\dagger } & 0 & \Omega _{22}%
\end{array}%
\right]
\end{equation}%
and $\Gamma \Gamma ^{\dagger }$ has the block-diagonal form 
\begin{equation}
\Gamma \Gamma ^{\dagger }=\left[ 
\begin{array}{cc}
\Gamma _{11}\Gamma _{11}^{\dagger } & 0 \\ 
0 & \Gamma _{22}\Gamma _{22}^{\dagger }%
\end{array}%
\right] ,
\end{equation}%
from which we see that any eigenvector of $\Gamma \Gamma ^{\dagger }$ is
decomposed with respect to $H_{1}=\textstyle{\bigoplus\limits_{\alpha =1}^{n}%
}H_{1\alpha }$ into a sum of eigenvectors (possibly zero) of $\Gamma \Gamma ^{\dagger }$ with the same eigenvalue. Thus each eigenspace of $\Gamma \Gamma ^{\dagger }$ admits an orthogonal decomposition into its intersections with all of the $%
H_{1\alpha }$. This is what allows us to choose the basis $\{g_{q}\}$ so
that each is contained in one of the $H_{1\alpha }$. It follows that $%
g_{q}^{\prime }$ is in $H_{2\alpha }$.

We then ask, given any choice of basis $\{g_{q}\}$, what is the finest
decomposition of the form \eqref{CalHdecomposition} such that each $g_{q}$
is contained in one of the $H_{1\alpha }$? The answer requires considering
the orbits of the vectors $g_{q}$ under $\Omega _{1}$ and the orbits of the
vectors $g_{q}^{\prime }$ under $\Omega _{2}$. We see that if $g_{q}\in
H_{1\alpha }$ for some $q$ and $\alpha $, it is required by the invariance
of $H_{i\alpha }$ under $\Omega _{i}$, for $i=1,2$, that $\mathcal{O}%
_{\Omega _{1}}(\{g_{q}\})\in H_{1\alpha }$ and $\mathcal{O}_{\Omega
_{2}}(\{g_{q}^{\prime }\})\in H_{2\alpha }$. Further, the orbit $\mathcal{O}%
_{\Omega _{1}}(\{g_{q}\})$ must be orthogonal to every $g_{p}$ that is not
in $H_{1\alpha }$ and the orbit $\mathcal{O}_{\Omega _{2}}(\{g_{q}^{\prime
}\})$ must be orthogonal to every $g_{p}^{\prime }$ that is not in $%
H_{2\alpha }$.

\begin{theorem}[canonical decomposition]
\label{thmDecomposition} Assume that the spectrum of $\Gamma \Gamma ^{\dag }$
consists of a countable set of eigenvalues (and their accumulation points).

\begin{enumerate}
\item Let an s-invariant splitting of $(\mathcal{H},\Omega)$ be given: 
\begin{equation}  \label{CalHdecomposition2}
\mathcal{H} = \textstyle{\bigoplus\limits_{\alpha=1}^n} \left( H_{1\alpha}
\oplus H_{2\alpha} \right).
\end{equation}
Then there exists an orthonormal Hilbert-space basis $\{g_q\}_{q=1}^r$ for 
$\mathrm{Ran}\,\Gamma$ of eigenvectors of $\Gamma\Gamma^\dagger$ and
corresponding basis $\{g^{\prime }_q = Ug_q\}$ for $\mathrm{Ran}\,\Gamma^\dagger$ such that for each $q$, there exists $\alpha$ such that $g_q\in H_{1\alpha}$ and
$g'_q\in H_{2\alpha}$.

\item Given an arbitrary choice of basis $\{g_{q}\}_{q=1}^{r}$ for $\mathrm{Ran}\,\Gamma $ consisting of eigenvectors of $\Gamma \Gamma ^{\dagger }$, it follows that 
the finest s-invariant splitting (of the form \eqref{CalHdecomposition2}) such that each $g_{q}$ is in some $H_{1\alpha }$ is obtained from the orbits
\begin{equation}
H_{1\alpha }=\mathcal{O}_{\Omega _{1}}(\{g_{q}:q\in V_{\alpha }\})\quad 
\text{\textrm{and}}\quad H_{2\alpha }=\mathcal{O}_{\Omega
_{2}}(\{g_{q}^{\prime }:q\in V_{\alpha }\}),
\end{equation}
in which the $V_{\alpha }$ are the minimal (disjoint) subsets of $\{1,\dots ,r\}$ such
that $\mathcal{O}_{\Omega _{1}}(\{g_{q}:q\in V_{\alpha }\})\perp g_{p}$ and $%
\mathcal{O}_{\Omega _{2}}(\{g_{q}^{\prime }:q\in V_{\alpha }\})\perp
g_{p}^{\prime }$ for all $p\not\in V_{\alpha }$.
\end{enumerate}
\end{theorem}

\noindent In part ({\itshape ii}) it is tacitly implied that such minimal
subsets are well defined.

\section{Proofs of theorems}

\label{sectionProofs}

In this section we formulate detailed statements on the stucture of
open systems and provide their proofs, which encompass the proofs of the theorems
from Sections \ref{sectionReconstructibility} and \ref{sectionDecomposition}. The development follows that of those sections.

We use the same notation as in the previous sections. We are given a
conservative system $(\mathcal{H},\Omega )$: 
\begin{equation}
\partial _{t}\mathcal{V}=-\mathrm{i}\Omega \mathcal{V+F},
\end{equation}%
and an orthogonal splitting of the Hilbert space 
\begin{equation}
\mathcal{H}=H_{1}\oplus H_{2},
\end{equation}%
with respect to which $\Omega $ has the form 
\begin{equation}
\Omega =\left[ 
\begin{array}{cc}
\Omega _{1} & \Gamma \\ 
\Gamma ^{\dagger } & \Omega _{2}%
\end{array}%
\right] ,
\end{equation}
and we let $P_1$ denote projection onto $H_1$ and define, as before, 
\begin{equation}
\mathring\Omega = \left[ 
\begin{array}{cc}
\Omega_1 & 0 \\ 
0 & \Omega_2%
\end{array}
\right], \quad \mathring\Gamma = \left[ 
\begin{array}{cc}
0 & \Gamma \\ 
\Gamma^\dagger & 0%
\end{array}
\right], \quad \Omega = \mathring\Omega + \mathring\Gamma,
\end{equation}
As $\mathrm{Ran}\,\Gamma $ and $\mathrm{Ran}\,\Gamma ^{\dag }$ are
isomorphic through the isomorphism $U$ \eqref{U}, we may let $H_{0}$ be
a standard Hilbert space on which the operator $\Gamma _{R}$ is represented
by a self-adjoint operator $\Gamma _{0}$. This means that there are unitary
operators 
\begin{equation}
U_{1}:H_{0}\rightarrow \mathrm{Ran}\,\Gamma \subseteq H_{1},\qquad
U_{2}:H_{0}\rightarrow \mathrm{Ran}\,\Gamma ^{\dagger }\subseteq H_{2}
\end{equation}%
with 
\begin{equation}
U=U_2U_{1}^{-1}:\mathrm{Ran}\,\Gamma \rightarrow \mathrm{Ran}\,\Gamma
^{\dagger }
\end{equation}%
such that 
\begin{equation}
\Gamma _{0}=U_{2}^{-1}\Gamma ^{\dagger }U_{1}=U_{1}^{-1}\Gamma U_{2}.
\end{equation}%
Since $\mathrm{Null}\,\Gamma ^{\dagger }\perp \mathrm{Ran}\,\Gamma $, $%
\Gamma ^{\dagger }$ is completely determined by its action on $\mathrm{Ran}%
\,\Gamma $, and the positive operator $\Gamma _{R}\Gamma _{R}^{\dagger }$
restricted to $\mathrm{Ran}\,\Gamma $ is represented by $\Gamma _{0}^{2}$ on 
$H_{0}$ through the isometric isomorphism given by $U_{1}$; the analogous
structure holds for $\Gamma _{R}^{\dagger }\Gamma _{R}$: 
\begin{equation}
\Gamma _{R}\Gamma _{R}^{\dagger }=U_{1}\Gamma _{0}^{2}U_{1}^{-1},\qquad
\Gamma _{R}^{\dagger }\Gamma _{R}=U_{2}\Gamma _{0}^{2}U_{2}^{-1}.
\end{equation}


\subsection{Reconstructibility from open subsystems}

\label{subsectionReconstructibility}

The following statement is a detailed version of Theorem \ref%
{thmReconstructibility}.

\begin{theorem}[system reconstructiblity]
\label{thmReconstructibility2} There exists a unique minimal sub-Hilbert-space
$\mathcal{H}^{\prime }$ of $\mathcal{H}$ with the following properties:

\renewcommand{\labelenumi}{\alph{enumi}.}

\begin{enumerate}
\item $\mathcal{H}^{\prime }$ is $\Omega$-invariant
($\orbit{\Omega}(\mathcal H') = \mathcal H'$)
and, hence, $(\mathcal{H}^{\prime },\Omega \!\!\upharpoonright \!\!{\mathcal{H}^{\prime }})$ is
conservative;

\item $\mathcal{H}^{\prime }$ is $P_{1}$-invariant, that is, $\mathcal{H}%
^{\prime }=H_{1c}\oplus H_{2c}$, where $H_{1c}\subseteq H_{1}$ and $%
H_{2c}\subseteq H_{2}$;

\item $\mathrm{Ran}\,\Gamma \subseteq \mathcal{H}^{\prime }$.
\end{enumerate}

\renewcommand{\labelenumi}{\roman{enumi}.}

\noindent Let \thinspace\ $\Omega _{ic}=\Omega _{i}\!\!\upharpoonright
\!\!H_{ic}$ for $i=1,2$,\thinspace\ $\Gamma _{c}=\Gamma
\!\!\upharpoonright \!\!H_{2c}$,\thinspace\ $a_{1}(t)=\Gamma \mathrm{e}^{-%
\mathrm{i}\Omega _{2}t}\Gamma ^{\dagger }$,\thinspace\ and $a_{2}(t)=\Gamma
^{\dagger }\mathrm{e}^{-\mathrm{i}\Omega _{1}t}\Gamma $. The the following
hold

\begin{enumerate}
\item $H_{1c} = \mathcal{O}_{\Omega_1}(\mathrm{Ran}\,\Gamma)$ and $H_{2c} = 
\mathcal{O}_{\Omega_2}(\mathrm{Ran}\,\Gamma^\dagger)$.

\item $(\mathcal{H}^{\prime }\!=\!H_{1c}\oplus
H_{2c},\,\Omega\!\!\upharpoonright\!\!{\mathcal{H}^{\prime }})$ is
reconstructible, \newline
$H_1\oplus H_{2c}$ is the unique minimal conservative extension of $%
(H_1,\Omega_1,a_1(t))$ contained in $\mathcal{H}$, and $H_{1c}\oplus H_2$ is
the unique minimal conservative extension of $(H_2,\Omega_2,a_2(t))$
contained in $\mathcal{H}$.

\item $a_{1}(t)\!\!\upharpoonright \!\!H_{1c}=\Gamma _{c}\,\mathrm{e}^{-%
\mathrm{i}\Omega _{2c}t}\,\Gamma _{c}^{\dagger }$ \thinspace and\thinspace\ $%
a_{1}(t)\!\!\upharpoonright \!\!(H_{1}\ominus H_{1c})=0$, \newline
$a_{2}(t)\!\!\upharpoonright \!\!H_{2c}=\Gamma _{c}^{\dagger }\,\mathrm{e}%
^{-\mathrm{i}\Omega _{1c}t}\,\Gamma _{c}$ \thinspace and\thinspace\ $%
a_{2}(t)\!\!\upharpoonright \!\!(H_{2}\ominus H_{2c})=0$. \newline
Thus the system $H_{1c}\oplus H_{2c}$ completely determines the friction
functions $a_{1}(t)$ and $a_{2}(t)$. Neither $a_{1}(t)$ nor $a_{2}(t)$ is
determined by any proper $\Omega $-invariant subsystem of $H_{1c}\oplus
H_{2c}$ of the form $\tilde{H}_{1}\oplus \tilde{H}_{2}$, where $\tilde{H}%
_{i}\subset H_{i}$, $i=1,2$.

\item $(H_{1}\!\oplus\! H_{2},\Omega)$ is reconstructible if and only if $H_1 $ and $H_2$ have no nontrivial $\Omega$-invariant subspaces.
\end{enumerate}
\end{theorem}

\begin{proof}
\noindent
There exists a sub-Hilbert-space $\mathcal H'$ of $\mathcal H$ possessing properties (a-c) because $\mathcal H$ is such a subspace.  Let $\mathcal H'$ be an arbitrary such space.
First, we show that $\mathrm{Ran}\,\Gamma ^{\dagger }\subset 
\mathcal{H}^{\prime }$. We have 
\begin{equation}
\mathrm{Ran}\,\Gamma ^{\dagger }=\Gamma ^{\dagger }(H_{1})=\Gamma ^{\dagger
}(\mathrm{Ran}\,\Gamma )
\end{equation}%
since $\mathrm{Null}\,\Gamma ^{\dagger }=H_{1}\ominus \overline{\mathrm{Ran}%
\,\Gamma }$.
Let $w\in \Gamma ^{\dagger }(\mathrm{Ran}\,\Gamma )$, say $w=\Gamma^\dag(u)$ for some $u\in \range(\Gamma) \subset \mathcal H' \cap H_1$.  Let $\epsilon>0$ be given.  Since $\Omega_1$ is densely defined in $H_1 \cap \mathcal H'$, there exists
$v\in\domain\Omega_1 \cap \mathcal H'$ such that
$\| u-v \| < \epsilon/\|\Gamma^\dag\|$ and hence
$\| w - \Gamma^\dag(v) \| = \| \Gamma^\dag(u-v) \| < \epsilon$.
By (a,b), we obtain
\begin{equation}
\Gamma ^{\dagger }(v)=(I-P_{1})(\Omega _{1}(v)+\Gamma ^{\dagger
}(v))=(I-P_{1})\Omega (v)\in \mathcal{H}^{\prime }.
\end{equation}%
Since $\epsilon$ is arbitrary, we conclude that $w\in \mathcal{H}^{\prime }$. We now know that 
\begin{equation}
{\mathcal{O}}_{\Omega }(\mathrm{Ran}\,\mathring{\Gamma})\subseteq \mathcal{H}%
^{\prime }.
\end{equation}
But, using the definition \eqref{dorbit2} of the orbit under the action of two operators,  we see that ${\mathcal{O}}_{\Omega }(\mathrm{Ran}\,\mathring{\Gamma})$
is itself $P_{1}$-invariant as follows: 
\begin{gather}
{\mathcal{O}}_{\Omega }(\mathrm{Ran}\,\mathring{\Gamma})={\mathcal{O}}%
_{\Omega ,\mathring{\Gamma}}(\mathrm{Ran}\,\mathring{\Gamma})={\mathcal{O}}_{%
\mathring{\Omega},\mathring{\Gamma}}(\mathrm{Ran}\,\mathring{\Gamma})={%
\mathcal{O}}_{\mathring{\Omega}}(\mathrm{Ran}\,\mathring{\Gamma}) \\
={\mathcal{O}}_{\mathring{\Omega}}(\mathrm{Ran}\,\Gamma \oplus \mathrm{Ran}%
\,\Gamma ^{\dagger })={\mathcal{O}}_{\Omega _{1}}(\mathrm{Ran}\,\Gamma
)\oplus {\mathcal{O}}_{\Omega _{2}}(\mathrm{Ran}\,\Gamma ^{\dagger }).
\end{gather}

Defining $H_{1c}$ and $H_{2c}$ as in (i), we see that $H_{1c}\oplus H_{2c}$ both satisfies properties (a-c) and is contained in our arbitrarily chosen $\mathcal H'$ with these properties.
This proves the uniqueness of a minimal subspace satisfying (a-c), namely,
$\mathcal H' = \orbit{\Omega}(\range\mathring\Gamma) = H_{1c}\oplus H_{2c}$, as well as property (i).

To prove that $(\mathcal{H}^{\prime },\,\Omega\!\!\upharpoonright\!\!{%
\mathcal{H}^{\prime }})$ is reconstructible (part (ii)), we must show that
$\mathcal H' = \orbit{\Omega}(H_{1c})$, which is the state space of the minimal conservative extension for $H_{1c}$ in $\mathcal{H}$, and that
$\mathcal H' = \orbit{\Omega}(H_{2c})$.  Observe that $\range\Gamma\in\mathcal{H}^{\prime }$ and choose again $w$ and $v$ as before. We have $\Omega_1(v) \in H_{1c}$, so that 
\begin{equation}
\Gamma^\dagger(v) = \Omega(v) - \Omega_1(v) \in {\mathcal{O}}%
_{\Omega}(H_{1c}),
\end{equation}
and $\| \Gamma^\dag(v) -w \| < \epsilon$.  We conclude that 
\begin{equation}
\mathrm{Ran}\,\mathring\Gamma \subseteq {\mathcal{O}}_{\Omega}(H_{1c}).
\end{equation}
But since $\mathcal{H}^{\prime }= {\mathcal{O}}_{\Omega}(\mathrm{Ran}%
\,\mathring\Gamma)$ and $H_{1c}\subseteq \mathcal{H}^{\prime }$, we obtain 
\begin{equation}
{\mathcal{O}}_{\Omega}(H_{1c}) = \mathcal{H}^{\prime }.
\end{equation}
An analogous argument applies to $H_{2c}$.

To prove the rest of part (ii), 
define $H_{1d} = H_1\ominus H_{1c}$ and $H_{2d} = H_2\ominus H_{2c}$. Since $%
H_{1d}\oplus H_{2d} = \mathcal{H }\ominus \mathcal{H}^{\prime }$, $%
H_{1d}\oplus H_{2d}$ is $\Omega$-invariant.  Since $H_{1d}$ is
perpendicular to $\mathrm{Ran}\,\Gamma$, for $v\in H_{1d}\cap\range{\Omega}$,
$\Omega(v) = \Omega_1(v) + \Gamma^\dag(v) = \Omega_1(v) \in H_{1d}$, and we see that $H_{1d}$ itself is 
$\Omega$-invariant. Therefore, so is $H_1\oplus H_{2c}$, and we obtain 
\begin{equation}
{\mathcal{O}}_{\Omega}(H_1) = \orbit{\Omega}(H_{1d}) \oplus \orbit{\Omega}(H_{1c}) = H_1\oplus H_{2c}
\end{equation}
That ${\mathcal{O}}%
_{\Omega}(H_2) = H_2\oplus H_{1d}$ is shown similarly.

To prove part (iii), let $v\in H_{1c}$ be given. Then 
\begin{eqnarray*}
a_{1}(t)v &=&\Gamma \mathrm{e}^{-\mathrm{i}\Omega _{2}t}\Gamma ^{\dagger
}v=\Gamma \mathrm{e}^{-\mathrm{i}\Omega _{2}t}\Gamma _{c}^{\dagger }v\qquad 
\text{\textrm{(because }}v\in H_{1c}\text{\textrm{)}} \\
&=&\Gamma \mathrm{e}^{-\mathrm{i}\Omega _{2c}t}\Gamma _{c}^{\dagger }v\qquad 
\text{\textrm{(because }}\Gamma _{c}^{\dagger }v\in \mathrm{Ran}\,\Gamma
^{\dagger }\subset H_{2c}\text{\textrm{)}} \\
&=&\Gamma _{c}\mathrm{e}^{-\mathrm{i}\Omega _{2c}t}\Gamma _{c}^{\dagger
}v\qquad \text{\textrm{(because }}\mathrm{e}^{-\mathrm{i}\Omega
_{2c}t}\Gamma _{c}^{\dagger }v\in H_{2c}\text{\textrm{)}}.
\end{eqnarray*}%
Let $v\in H_{1}\ominus H_{1c}$ be given. Then $v\perp \mathrm{Ran}\,\Gamma $, so that $v\in \mathrm{Ker}\,\Gamma ^{\dagger }$. The analogous statement
about $a_{2}(t)$ is proven similarly.
Finally, if $\tilde H_1\oplus\tilde H_2$ is a proper subspace of $\mathcal H' = H_{1c}\oplus H_{2c}$ that is invariant under $\Omega$ and $P_1$, then
$\mathrm{Ran}\,\Gamma\not\subset\tilde H_1$.  This is because $\mathcal H'$ is the minimal such space that contains $\range\Gamma$.
Since $a_1(0) = \Gamma\Gamma^\dagger$, we have $\mathrm{Ran}\, a_1(0) = 
\mathrm{Ran}\,\Gamma \not\subset\tilde H_1$.  This means that
the restriction of the system $(\mathcal H',\Omega\restrict\mathcal H')$
to $\tilde H_1\oplus\tilde H_2$ does not determine $a_1(0)$, and therefore does not determine the function $a_1(t)$.

To prove part (\textit{iv}), first observe that, if $H_1\oplus H_2$ is not
reconstructible, then $H_{1d}$, which is $\Omega$-invariant, is nontrivial.
Conversely, suppose that $H_1$ has an $\Omega$-invariant subspace $%
H_1^{\prime }\subset H_1$. Set $H_1^{\prime \prime }= H_1\ominus H_1^{\prime
}$, and let $H_1^{\prime \prime }\oplus H_2^{\prime \prime }= {\mathcal{O}}%
_{\Omega}(H_1^{\prime \prime })$. Then the minimal conservative extension of 
$(H_1,\Omega_1,a_1(t))$ is ${\mathcal{O}}_{\Omega}(H_1) = H_1^{\prime }\oplus
H_1^{\prime \prime }\oplus H_2^{\prime \prime }$, so that $H_2^{\prime
\prime }= H_{2c}$. For all $v\in H_{2c}\cap\mathrm{Dom}\,(\Omega)$, $%
\Omega(v) = \Gamma(v) + \Omega_2(v) \in H_1^{\prime \prime }\oplus H_{2c}$.
Thus $\Gamma(v) \perp H_1^{\prime }$, so that $\mathrm{Ran}\,\Gamma \perp
H_1^{\prime }$ and hence $\mathrm{Ran}\,\Gamma \subset H_1^{\prime
\prime }$. Since $H_1^{\prime \prime }\oplus H_{2c} = {\mathcal{O}}%
_{\Omega}(H_1^{\prime \prime }\oplus H_{2c}) \subset {\mathcal{O}}_{\Omega}(%
\mathrm{Ran}\,\mathring\Gamma) = H_{1c}\oplus H_{2c}$, we obtain $H_1^{\prime}\perp H_{1c}$ so that $H_1\oplus H_2$ is not reconstructible.
\end{proof}

\begin{remark}
Of all $\Omega $-invariant subsystems of the form $\tilde{H}_{1}\oplus 
\tilde{H}_{2}$ with $\tilde{H}_{i}\subset H_{i}$, $i=1,2$, $H_{1c}\oplus
H_{2c}$ is the minimal reconstructible one that has the property that $%
\mathrm{Ran}\,\Gamma \subset H_{1c}$. There may exist $\Omega $-invariant
subsystems of the same form such that $\mathrm{Ran}\,\Gamma \not\subset
H_{1c}$ (this is dealt with in Theorem \ref{thmDecomposition}) and
$\Omega$-invariant subsystems that are not of this form, which do not concern us.
\end{remark}

Theorems \ref{thmSpectrum}, \ref{thmSpectrum2}, and \ref{thmMultiplicity} are rather straightforward applications of standard spectral
theory of self-adjoint operators in separable Hilbert space. We shall set
down the general framework and prove those results. The relevant
material can be found, for example, in Akhiezer and Glazman \cite{AkhGlaz}
or \cite{RiNa}.

Let $dE_\lambda$ be the spectral resolution of the identity for a
self-adoint operator $\Omega$ in the Hilbert space $\mathcal{H}$. This means
that $dE_\lambda$ is an (orthogonal) projection-valued Borel measure on $%
\mathbb{R}$ such that, for each $v\in\mathcal{H}$, the vector-valued
function of $\mu$ given by 
\begin{equation}
\int_{(-\infty,\mu]} \hspace{-10pt}  d(E_\lambda v)
\end{equation}
is right-continuous, 
\begin{equation}
\int_{\mathbb{R}} d(E_\lambda v) = \lim_{\mu\to\infty} \int_{(-\infty,\mu]} 
\hspace{-10pt} d(E_\lambda v) = v,
\end{equation}
and, for each $f\in C_c(\mathbb{R})$, 
\begin{equation}
f(\Omega) v = \int_{\mathbb{R}} f(\lambda) \, d(E_\lambda v)\, ;
\end{equation}
integration is understood in the Lebesgue-Stieltjes sense.

The orbit of a subset $S\subset \mathcal{H}$ generated by the action of $%
\Omega$ can be expressed in terms of continuous functions of $\Omega$ or in
terms of spectral projections: 
\begin{eqnarray}
{\mathcal{O}}_{\Omega}(S) &=& \text{closure of span} \left\{ f(\Omega) v :
v\in S, f\in C_c(\mathbb{R}) \right\} \\
&=& \text{closure of span} \left\{ \int_B  d(E_\lambda v) : v\in S \text{, $%
B $ a Borel set} \right\}  \label{orbit1} \\
&=& \text{closure of span} \left\{ \int_\Delta d(E_\lambda v) : v\in S \text{%
, $\Delta$ an interval in $\mathbb{R}$ with $|\Delta|<\epsilon$} \right\},
\label{orbit2}
\end{eqnarray}
in which $\epsilon$ is an arbitrary positive real number.

\smallskip

\begin{proof}[Proof of Theorems \protect\ref{thmSpectrum}, \protect\ref{thmSpectrum2}, and \protect\ref{thmMultiplicity}]
Parts (\textit{ii}) of Theorems \ref{thmSpectrum} and \ref{thmSpectrum2} are
statements of the representation \eqref{orbit2} of the $\Omega$-orbits of $%
H_1$ and $\mathrm{Ran}\,\mathring\Gamma$ and the $\Omega_2$-orbit of $%
\mathrm{Ran}\,\Gamma^\dag$. We have already shown (see equation \ref%
{OrbitGamma}) that $\mathcal{H}_\text{{\scriptsize {min}}}$ is generated
through the action of $\Omega$ on the range of $\mathring\Gamma$.

To prove parts (\textit{i}) of these theorems, observe that, in the case of
pure point spectrum, 
\begin{equation}
dE_\lambda = \sum_{j=1}^N E_j \delta(\lambda_j - \lambda),
\end{equation}
in which $\delta$ is the unit measure concentrated at $\lambda=0$, the $\lambda_j$ are the distinct eigenvalues of $\Omega$, the $E_j$ are orthogonal projections, and $N$ may be equal to $\infty$. Consider the representation \eqref{orbit1}: for any vector $v\in S$ and Borel set $B$, 
\begin{equation}
\int_B d(E_\lambda v) = \sum_{j:\lambda_j\in B} E_j v.
\end{equation}
In particular, $E_j v \in {\mathcal{O}}_{\Omega}(S)$ (by taking $B=\{
\lambda_j \}$) for $j=1,\dots,N$, and each vector $\int_B d(E_\lambda v)$ is
in the closure of the linear span of the $E_j v$. It follows that 
\begin{equation}
{\mathcal{O}}_{\Omega}(S) = \text{closure of span} \left\{ E_j v : j\in
\{1,\dots,N\}, \, v\in S \right\} = \bigoplus_{i=1}^N E_j(S).
\end{equation}
The statements (\textit{i}) of the theorems follow from applying this result
to the respective operator $\Omega$ and set $S$.

To prove parts (\textit{iii}) of the theorems, observe that, since $\mathcal{%
H }= \bigoplus_{i=1}^N \mathrm{Ran}\,(E_j)$, 
\begin{equation}
{\mathcal{O}}_{\Omega}(S)^\perp = \bigoplus_{i=1}^N \left( \mathrm{Ran}%
\,(E_j) \ominus E_j(S) \right).
\end{equation}
Suppose that $S$ is a subspace of $\mathcal{H}$, and let $P$ denote the
projection onto $S$. Then, for any eigenvector, say $\phi\in \mathrm{Ran}%
\,(E_j)$, we have a splitting $\phi = \phi_1 + \phi_2$, where $\phi_1\in
E_j(S)\subseteq {\mathcal{O}}_{\Omega}(S)$ and $\phi_2\in \mathrm{Ran}%
\,(E_j) \ominus E_j(S) \subseteq {\mathcal{O}}_{\Omega}(S)^\perp$. Thus, $%
P(\phi_1)=0$ and $\phi_2 = E_j(v)$ for some $v\in S$. From 
\begin{equation}
\| \phi_2 \|^2 = \| E_j(v) \|^2 = \langle v | E_j(v) \rangle = \langle v |
\phi_2 \rangle,
\end{equation}
we infer that $\phi_2=0$ if and only if $\phi_2\perp S$, that is, if and
only if $P(\phi_2) = 0$. But we also see that $\phi_2=0$ if and only if $%
\phi\perp E_j(S)$, which is true if and only if $\phi\in{\mathcal{O}}%
_{\Omega}(S)^\perp$. It follows that 
\begin{equation}  \label{yow}
{\mathcal{O}}_{\Omega}(S)^\perp = \{0\} \text{ if and only if $P(\phi)\not=0$
for each (nonzero) eigenvector of $\Omega$}.
\end{equation}

This result applies directly to the first part of (\textit{iii}) of Theorem %
\ref{thmSpectrum} because $\mathcal{H}_\text{{\scriptsize {min}}} = {%
\mathcal{O}}_{\Omega}(H_1)$. For the second part and part (\textit{iii}) of
Theorem \ref{thmSpectrum2}, \eqref{yow} applies after observing that (1) $%
\mathcal{H}_\text{{\scriptsize {min}}} = H_1\oplus H_{2c} = H_{1c} \oplus
H_{2c} = {\mathcal{O}}_{\Omega}(\mathrm{Ran}\,\mathring\Gamma)$ if and only
if $H_{1c}=H_1$ and (2) $P$ denotes projection onto $\mathrm{Ran}%
\,\mathring\Gamma$, then for each $v\in\mathcal{H}$, $P(v) = 0$ if and only
if $\Gamma(v) = 0$---this is because $\mathring\Gamma$ is self-adjoint so
that the nullspace of $\mathring\Gamma$ is equal to $(\mathrm{Ran}%
\,\mathring\Gamma)^\perp$.

Theorem \ref{thmMultiplicity} follows from the fact that, if $S$ is a
subspace of $\mathcal{H}$, then the multiplicity of $\Omega$ restricted to ${%
\mathcal{O}}_{\Omega}(S)$ is bounded by the dimension of $S$.
\end{proof}


\subsection{Decomposition of coupled systems}

\label{subsectionDecomposition}


The next statement on the equivalence between decoupling of an open system
and s-invariant decompositions of its minimal conservative extension is a detailed
version of Theorem \ref{thmOpen}.

\begin{theorem}[decoupling and s-invariant decomposition]
\label{thmOpen2} Let an open linear system $(H_{1},\Omega _{1},a_{1}(t))$ be
given, and let $(\mathcal{H},\Omega )$ be its minimal conservative
extension, with $P_{1}$, $H_{2}=\mathcal{H}\ominus H_{1}$, $\Omega _{2}$,
and $\Gamma :H_{2}\rightarrow H_{1}$ defined as before.

\begin{enumerate}
\item Let $\mathcal{H}=\mathcal{H}^{\prime }\oplus \mathcal{H}^{\prime
\prime }$ be an s-invariant decomposition, with $H_{1}=H_{1}^{\prime }\oplus
H_{1}^{\prime \prime }$, where $H_{1}^{\prime }=P_{1}(\mathcal{H}^{\prime })$%
. Then the open system $(H_{1},\Omega _{1},a_{1}(t))$ is decoupled, that is,
if $\pi _{1}^{\prime }$ and $\pi _{1}^{\prime \prime }$ are projections in $%
H_{1}$ onto $H_{1}^{\prime }$ and $H_{1}^{\prime \prime }$, then $\pi
_{1}^{\prime }a_{1}(t)\pi _{1}^{\prime \prime }=0$ and $\pi _{1}^{\prime
\prime }a_{1}(t)\pi _{1}^{\prime }=0$. Equivalently, putting $v_{1}^{\prime
}(t)=\pi _{1}^{\prime }v(t)$, $a_{1}^{\prime }(t)=\pi _{1}^{\prime
}a_{1}(t)\pi _{1}^{\prime }$ and $f_{1}^{\prime }(t)=\pi _{1}^{\prime }f(t)$%
, the dynamics of the open system $(H_{1},\Omega _{1},a_{1}(t))$ are
decoupled into 
\begin{equation}
\partial _{t}v_{1}^{\prime }(t)=-\mathrm{i}\Omega _{1}v_{1}^{\prime
}(t)-\int_{0}^{\infty }a_{1}^{\prime }(\tau )v_{1}^{\prime }(t-\tau )\,%
\mathrm{d}\tau +f_{1}^{\prime }(t)  \label{open1'}
\end{equation}%
and 
\begin{equation}
\partial _{t}v_{1}^{\prime \prime }(t)=-\mathrm{i}\Omega _{1}v_{1}^{\prime
\prime }(t)-\int_{0}^{\infty }a_{1}^{\prime \prime }(\tau )v_{1}^{\prime
\prime }(t-\tau )\,\mathrm{d}\tau +f_{1}^{\prime \prime }(t).
\label{open1''}
\end{equation}

\item Let $H_1 = H^{\prime }_1 \oplus H^{\prime \prime }_1$ be an $\Omega_1$%
-invariant decomposition with corresponding projections $\pi^{\prime }_1$
and $\pi^{\prime \prime }_1$, and suppose that $\pi^{\prime
}_1a_1(t)\pi^{\prime \prime }_1=0$, that is, that \eqref{open1'} holds (the
evolution of $H^{\prime }_1$ is not influenced by $H^{\prime \prime }_1$).
Then there exists an s-invariant decomposition $\mathcal{H }= \mathcal{H}%
^{\prime }\oplus \mathcal{H}^{\prime \prime }$, with projection $\pi^{\prime
}$ onto $\mathcal{H}^{\prime }$, such that $H^{\prime }_1 = \mathrm{Ran}\,
\pi^{\prime }P_1$. In addition, $(\mathcal{H}^{\prime
},\Omega\!\!\upharpoonright\!\!{\mathcal{H}^{\prime }})$ is the minimal
conservative extension of $(H^{\prime }_1,\Omega_1\!\!\upharpoonright\!\!{%
H^{\prime }_1},\pi^{\prime }_1a(t)\pi^{\prime }_1)$.
\end{enumerate}
\end{theorem}

\begin{proof}
The proof of the first part is straightforward. To prove the second
statement, let 
\begin{equation}
a_{1}^{\prime }(t)=\pi _{1}^{\prime }a_{1}(t)\pi _{1}^{\prime },\quad
a_{1}^{\prime \prime }(t)=\pi _{1}^{\prime \prime }a_{1}(t)\pi _{1}^{\prime
\prime },\quad \tilde{a}_{1}^{\prime \prime }(t)=\pi _{1}^{\prime \prime
}a_{1}(t)\pi _{1}^{\prime },
\end{equation}%
so that, in block-matrix form with respect to the decomposition $%
H_{1}=H_{1}^{\prime }\oplus H_{1}^{\prime \prime }$, $a_{1}(t)$ has the
representation 
\begin{equation}
a_{1}(t)=\left[ 
\begin{array}{cc}
a_{1}^{\prime }(t) & 0 \\ 
\tilde{a}_{1}^{\prime \prime }(t) & a_{1}^{\prime \prime }(t)%
\end{array}%
\right] .
\end{equation}%
Since $a_{1}(0)=\Gamma \Gamma ^{\dagger }$ is self-adjoint, we have 
\begin{equation}
\tilde{a}_{1}^{\prime \prime }(0)=\pi _{1}^{\prime \prime }a_{1}(0)\pi
_{1}^{\prime }=(\pi _{1}^{\prime }a_{1}(0)\pi _{1}^{\prime \prime
})^{\dagger }=0,
\end{equation}%
or, in matrix form, 
\begin{equation}
\Gamma \Gamma ^{\dagger }=a_{1}(0)=\left[ 
\begin{array}{cc}
a_{1}^{\prime }(0) & 0 \\ 
0 & a_{1}^{\prime \prime }(0)%
\end{array}%
\right] .
\end{equation}%
This form gives rise to a splitting of $\mathrm{Ran}\,\Gamma $ that is
invariant under $\Gamma _{R}\Gamma _{R}^{\dagger }$: 
\begin{equation}
\mathrm{Ran}\,\Gamma =\mathrm{Ran}\,a_{1}^{\prime }(0)\oplus \mathrm{Ran}%
\,a_{1}^{\prime \prime }(0)=\pi _{1}^{\prime }\mathrm{Ran}\,\Gamma \oplus
\pi _{1}^{\prime \prime }\mathrm{Ran}\,\Gamma =U_{1}(H_{0}^{\prime })\oplus
U_{1}(H_{0}^{\prime \prime }),
\end{equation}%
in which 
\begin{equation}
H_{0}=H_{0}^{\prime }\oplus H_{0}^{\prime \prime }=U_{1}^{-1}\mathrm{Ran}%
\,a_{1}^{\prime }(0)\oplus U_{1}^{-1}\mathrm{Ran}\,a_{1}^{\prime \prime }(0)
\end{equation}%
is the induced $\Gamma _{0}$-invariant splitting of the standard Hilbert
space $H_{0}$ for $\Gamma $ (by virtue of $\Gamma _{R}\Gamma _{R}^{\dagger
}=U_{1}\Gamma _{0}^{2}U_{1}^{-1}$). This gives a $\Gamma _{R}^{\dagger
}\Gamma _{R}$-invariant splitting of $\mathrm{Ran}\,\Gamma ^{\dagger }$: 
\begin{equation}
\mathrm{Ran}\,\Gamma ^{\dagger }=U_{2}(H_{0}^{\prime })\oplus
U_{2}(H_{0}^{\prime \prime })=\Gamma ^{\dagger }(H_{1}^{\prime })\oplus
\Gamma ^{\dagger }(H_{1}^{\prime \prime }).
\end{equation}%
and ultimately a splitting of the action of $\Gamma $ on $\mathrm{Ran}%
\,\Gamma ^{\dagger }$ and the action of $\Gamma ^{\dagger }$ on $\mathrm{Ran}%
\,\Gamma $: 
\begin{align}
\Gamma _{R}:U_{2}(H_{0}^{\prime })\rightarrow U_{1}(H_{0}^{\prime }),& \quad
\Gamma _{R}:U_{2}(H_{0}^{\prime \prime })\rightarrow U_{1}(H_{0}^{\prime
\prime });  \label{gammadecomp} \\
\Gamma _{R}^{\dagger }:U_{1}(H_{0}^{\prime })\rightarrow U_{2}(H_{0}^{\prime
}),& \quad \Gamma _{R}^{\dagger }:U_{1}(H_{0}^{\prime \prime })\rightarrow
U_{2}(H_{0}^{\prime \prime }).  \notag
\end{align}

We now prove that $H_{2}$ is decomposed into the $\Omega _{2}$-orbits of $%
U_{2}(H_{0}^{\prime })$ and $U_{2}(H_{0}^{\prime \prime })$. First, $\mathrm{%
Ran}\,(\Gamma ^{\dagger }\pi _{1}^{\prime \prime })\!=\!U_{2}(H_{0}^{\prime
\prime })$, and since $\pi _{1}^{\prime }\Gamma \mathrm{e}^{-\mathrm{i}%
\Omega _{2}t}\Gamma ^{\dagger }\pi _{1}^{\prime \prime }\!=\!0$ for all $t$,
we have \linebreak $\pi _{1}^{\prime }\Gamma ({\mathcal{O}}_{\Omega
_{2}}(U_{2}(H_{0}^{\prime \prime })))=\{0\}$. It follows that ${\mathcal{O}}%
_{\Omega _{2}}(U_{2}(H_{0}^{\prime \prime }))$ is orthogonal to $%
U_{2}(H_{0}^{\prime })$. Setting 
\begin{equation}
H_{2}^{\prime }={\mathcal{O}}_{\Omega _{2}}(U_{2}(H_{0}^{\prime }))\quad 
\text{and}\quad H_{2}^{\prime \prime }={\mathcal{O}}_{\Omega
_{2}}(U_{2}(H_{0}^{\prime \prime })),
\end{equation}%
we have $H_{2}^{\prime }\perp H_{2}^{\prime \prime }$ and 
\begin{equation}
\mathrm{Ran}\,\Gamma ^{\dagger }\subset H_{2}^{\prime }\oplus H_{2}^{\prime
\prime }\subset H_{2}.
\end{equation}%
As $H_{2}^{\prime }\oplus H_{2}^{\prime \prime }$ is an $\Omega _{2}$%
-invariant subspace of $H_{2}$ and $(\mathcal{H}=H_{1}\oplus H_{2},\Omega )$
is minimal as a conservative extension of $(H_{1},\Omega _{1},a_{1}(t))$, we
obtain 
\begin{equation}
H_{2}=H_{2}^{\prime }\oplus H_{2}^{\prime \prime }.
\end{equation}%
By part ({\itshape i}) of Theorem \ref{thmReconstructibility2}, $(\mathcal{H}%
^{\prime },\Omega \!\!\upharpoonright \!\!{\mathcal{H}^{\prime }})$ is
minimal as a conservative extension of $(H_{1}^{\prime },\Omega
_{1}\!\!\upharpoonright \!\!{H_{1}^{\prime }},a_{1}^{\prime }(t))$.
\end{proof}



\vspace{1ex}

The following statement describes a relation between the coupling operator $%
\Gamma $ and s-invariant decompositions, for $\Gamma $ with pure point
spectrum as in Theorem \ref{thmDecomposition}.

\begin{theorem}[coupling operator and s-invariant decomposition]
\label{thmDecomposition2} Suppose that $\Gamma _{0}$ has pure point spectrum.

\begin{enumerate}
\item Let 
\begin{equation}
H_1 = \textstyle{\bigoplus\limits_{\alpha=1}^n H_{1\alpha}} \quad \text{and}
\quad H_2 = \textstyle{\bigoplus\limits_{\alpha=1}^n H_{2\alpha}}
\end{equation}
be orthogonal decompositions such that $H_{1\alpha}\oplus H_{2\alpha}$ is
invariant under $\Omega$ for each $\alpha$, that is, the system $(\mathcal{H}%
,\Omega)$ splits as 
\begin{equation}  \label{splitting}
\mathcal{H} = \textstyle{\bigoplus\limits_{\alpha=1}^n} \left( H_{1\alpha}
\oplus H_{2\alpha} \right).
\end{equation}
Then there exists a decomposition of $\Gamma_0$ into rank-one operators 
\begin{equation}  \label{Gamma0Decomposition}
\Gamma_0 = \sum_{q=1}^r \sqrt{\gamma _{q}} \left\vert g_{0q}\right\rangle
\left\langle g_{0q}\right\vert, \qquad \left\langle g_{0p}\right\vert
\left\vert g_{0q}\right\rangle = \left\langle g_{0p}\right\vert \left\vert
g_{0q}\right\rangle =\delta_{pq},
\end{equation}
giving rise to a decomposition of $\Gamma$ into rank-one operators: 
\begin{equation}  \label{GammaDecomposition}
\Gamma = \sum_{q=1}^r \sqrt{\gamma _{q}} \left\vert g_{q}\right\rangle
\left\langle g^{\prime }_q\right\vert, \qquad \left\langle g_{p}\right\vert
\left\vert g_{q}\right\rangle = \left\langle g_{p}^{\prime }\right\vert
\left\vert g_{q}^{\prime }\right\rangle =\delta_{pq},
\end{equation}
such that, for $q=1,\dots,r$, there exists $\alpha$ such that $g_q \in
H_{1\alpha}$ and $g^{\prime }_q \in H_{2\alpha}$.

\item Conversely, assume that $\mathcal{H}$ is reconstructible, and let a
decomposition \eqref{GammaDecomposition} of $\Gamma$ be given arbitrarily.
Let $G$ be the graph with vertex set $\{1,\dots,r\}$ having an edge between $%
p$ and $q$ if and only if one of the following holds: 
\begin{equation}  \label{EdgeConditions}
\mathcal{O}_{\Omega_1}(g_p)\not\perp g_q, \quad \mathcal{O}%
_{\Omega_2}(g^{\prime }_p)\not\perp g^{\prime }_q.
\end{equation}
Let $V_1,\dots V_n$ be the vertex sets of the connected components of $G$,
and put 
\begin{equation}
H_{1\alpha} = \mathcal{O}_{\Omega_1}\left( \{g_q:q\in V_\alpha\} \right),
\quad H_{2\alpha} = \mathcal{O}_{\Omega_2}\left( \{g^{\prime }_q:q\in
V_\alpha\} \right).
\end{equation}

\begin{enumerate}
\item $H_1 = \textstyle{\bigoplus_{\alpha=1}^n H_{1\alpha}}$ and $H_2 = %
\textstyle{\bigoplus_{\alpha=1}^n H_{2\alpha}}$ are orthogonal
decompositions, and $H_{1\alpha}\oplus~H_{2\alpha}$ is invariant under $%
\Omega$ for each $\alpha$.

\item $\mathcal{H} = \textstyle{\bigoplus_{\alpha=1}^n} \left( H_{1\alpha}
\oplus H_{2\alpha} \right)$ is the finest s-invariant decomposition with the
property that each $g_q$ is in one of the $H_{1\alpha}$. This means that, if 
$\mathcal{H} = \textstyle{\bigoplus_{\beta=1}^m} \left( H_1^\beta \oplus
H_2^\beta \right)$ is another such decomposition, then, for all $%
\alpha=1,\dots,n$, there is $\beta$ such that $H_{i\alpha}\subset H_i^\beta$%
, for $i=1,2$.
\end{enumerate}

\item If all of the eigenspaces of $\Gamma\Gamma^\dagger$ are of dimension
1, then there exists a unique finest s-invariant decomposition of $(\mathcal{%
H},\Omega)$ (see form \eqref{splitting}). This means that, if \ $\mathcal{H}%
\!=\!\textstyle{\bigoplus\limits_{\beta=1}^m} \left( H_1^\beta \oplus
H_2^\beta \right)$ is any other s-invariant decomposition of $(\mathcal{H}%
,\Omega)$, then for all $\alpha=1,\dots,n$, there exists $\beta$ such that $%
H_{i\alpha}\in H_i^\beta$.
\end{enumerate}
\end{theorem}

Observe that the conditions \eqref{EdgeConditions} are equivalent to 
\begin{equation}
{\mathcal{O}}_{\Omega_1}(g_p) \perp {\mathcal{O}}_{\Omega_1}(g_q), \quad {%
\mathcal{O}}_{\Omega_2}(g_p^{\prime }) \perp {\mathcal{O}}%
_{\Omega_2}(g_q^{\prime }).
\end{equation}

\begin{proof}
\begin{enumerate}
\item From the $\Omega$-invariance of $H_{1\alpha}\oplus H_{2\alpha}$ for
each $\alpha=1,\dots,n$, we infer that $\Gamma^\dagger(H_{1\alpha})\subset
H_{2\alpha}$ and $\Gamma(H_{2\alpha})\subset H_{1\alpha}$ for each $\alpha$.
Define 
\begin{equation}
\Gamma_\alpha := \Gamma\!\!\upharpoonright\!\!{H_{2\alpha}}:
H_{2\alpha}\to H_{1\alpha}.
\end{equation}
It is straightforward to verify that 
\begin{equation}
\Gamma_\alpha^\dagger = (\Gamma_\alpha)^\dagger =
\Gamma^\dagger\!\!\upharpoonright\!\!{H_{1\alpha}}: H_{1\alpha}\to
H_{2\alpha},
\end{equation}
and we obtain a decomposition of $\Gamma$: 
\begin{equation}
\Gamma = \sum_{\alpha=1}^n \Gamma_\alpha \pi_{2\alpha},
\end{equation}
where $\pi_{2\alpha}$ is the orthogonal projection to $H_{2\alpha}$. For
each $\alpha$, $\Gamma_\alpha$ admits a decomposition into rank-one
operators 
\begin{equation}
\Gamma_{\alpha} = \sum_{q=1}^{r_\alpha} \sqrt{\gamma_q} \left\vert g_{\alpha
q}\right\rangle \left\langle g^{\prime }_{\alpha q}\right\vert, \qquad
\left\langle g_{\alpha p}\right\vert \left\vert g_{\alpha q}\right\rangle =
\left\langle g^{\prime }_{\alpha p}\right\vert \left\vert g^{\prime
}_{\alpha q}\right\rangle =\delta_{pq},
\end{equation}
where $\gamma_{\alpha q}>0$ are the eigenvalues of $\Gamma_\alpha\Gamma_%
\alpha^\dagger$ with corresponding orthonormal eigenvector basis $%
\{g_{\alpha q}\}_{q=1}^{r_\alpha}$ for $H_{1\alpha}$ and $\{g^{\prime
}_{\alpha q}\}_{q=1}^{r_\alpha}$ for $H_{2\alpha}$. Let $\{\gamma_q%
\}_{q=1}^r $ be an arrangement of $\{\gamma_{\alpha q}:
q=1,\dots,r_\alpha,\, \alpha=1,\dots,n\}$ and $\{g_q\}_{q=1}^r$ and $%
\{g^{\prime }_q\}_{q=1}^r$ the corresponding arrangements of the
eigenvectors $\{g_{\alpha q}\}$ and $\{g^{\prime }_{\alpha q}\}$. We obtain
the required form 
\begin{equation}
\Gamma = \sum_{\alpha=1}^n \Gamma_\alpha \pi_{2\alpha} = \sum_{q=1}^r \sqrt{%
\gamma _{q}} \left\vert g_{q}\right\rangle \left\langle g^{\prime
}_q\right\vert.
\end{equation}

\item 
\begin{enumerate}
\item Assume that $n>1$, and let $\alpha$ and $\beta$ be given with $%
1\leq\alpha,\beta\leq n$ and $\alpha\not=\beta$. Let $p\in V_\alpha$ and $%
q\in V_\beta$ be given. Since there is no edge in $G$ between $p$ and $q$,
we see that $\mathcal{O}_{\Omega_1}(g_p) \perp g_q$. $H_{1\alpha}$ is the
smallest $\Omega_1$-invariant subspace of $H_1$ containing $\{g_p: p\in
V_\alpha\}$, and is therefore equal to 
\begin{equation}
H_{1\alpha} = \sum\{\mathcal{O}_{\Omega_1}(g_p): p\in V_\alpha\}.
\end{equation}
We infer that $g_q\in H_1\ominus H_{1\alpha}$ for all $q\in V_\beta$. By the
self-adjointness of $\Omega_1$, $H_1\ominus H_{1\alpha}$ is $\Omega_1$%
-invariant, so that $H_{1\beta} = \mathcal{O}_{\Omega_1}(\{g_q: q\in
V_\beta\}) \subset H_1\ominus H_{1\alpha}$, and we conclude that $%
H_{1\alpha} \perp H_{1\beta}$. Now, $\mathcal{O}_{\Omega_1}(\mathrm{Ran}%
\,\Gamma) = \mathcal{O}_{\Omega_1}(\{g_q\}_{q=1}^r) = \bigoplus_{\alpha=1}^n
H_{1\alpha}$, and since $(\mathcal{H},\Omega)$ is reconstructible, $H_1 =
H_{1c} = \mathcal{O}_{\Omega_1}(\mathrm{Ran}\,\Gamma)$ by Theorem \ref%
{thmReconstructibility}. Therefore $H_1 = \bigoplus_{\alpha=1}^n H_{1\alpha}$%
. The analogous argument proves that $H_2 = \bigoplus_{\alpha=1}^n
H_{2\alpha}$. We now prove the invariance of $H_{1\alpha}\oplus H_{2\alpha}$
under $\Omega$. $H_{i\alpha}$ is by construction invariant under $\Omega_i$.
Let $v\in H_{1\alpha}$. Then $v=u+w$ for some $u\in\mathrm{span}\,\{g_q:
q\in V_\alpha\}$ and $w\in H_{1\alpha}\ominus \mathrm{span}\,\{g_q: q\in
V_\alpha\}$ Since $w \perp H_1\ominus H_{1\alpha}$, we see that $w \perp 
\mathrm{span}\,\{g_q: q\not\in V_\alpha\}$, so that $w \perp \mathrm{Ran}%
\,\Gamma$ and therefore $w\in\mathrm{Ker}\,\Gamma^\dagger$. We now obtain $%
\Gamma^\dagger(v) = \Gamma^\dagger(u) \in \mathrm{span}\,\{g^{\prime }_q:
q\in V_\alpha\} \subset H_{2\alpha}$. The invariance of $H_{1\alpha}\oplus
H_{2\alpha}$ under $\Omega$ now follows.

\item Let $\mathcal{H} = \bigoplus_{\beta=1}^m \left(H_1^\beta\oplus
H_2^\beta\right)$ be a $\Omega$-invariant decomposition of $\mathcal{H}$
such that each $g_q$ is in one of the $H_1^\beta$. Fix $\alpha$, and let $%
p,q\in V_\alpha$, so that $g_p,g_q\in H_{1\alpha}$ and $g^{\prime
}_q,g^{\prime }_p\in H_{2\alpha}$. Since $G$ contains an edge between $p$
and $q$, one of the conditions \eqref{EdgeConditions} is satisfied. Assume
that it is that $\mathcal{O}_{\Omega_1}(g_p) \not\perp g_q$ (the other case
is handled analogously). Let $\beta$ be such that $g_p \in H_1^\beta$; it
follows that $g^{\prime }_p\in H_2^\beta$. Since $\mathcal{O}%
_{\Omega_1}(g_p)\subset H_1^\beta$ and $H_1^\beta \perp H_1^{\beta^{\prime
}} $ for $\beta\not=\beta^{\prime }$, we have $g_q\in H_1^\beta$. We
conclude that $g_q\in H_1^\beta$ and $g^{\prime }_q\in H_2^\beta$ for all $%
q\in V_\alpha$, so that $H_{1\alpha}\subset H_1^\beta$ and $%
H_{2\alpha}\subset H_2^\beta$.
\end{enumerate}

\item Suppose the spectrum of $\Gamma _{0}$ is simple. Then the
representations \eqref{Gamma0Decomposition} and \eqref{Gamma0Decomposition}
are unique, so that by part (i), each decomposition $\bigoplus_{\beta
=1}^{m}\left( H_{1}^{\beta }\oplus H_{2}^{\beta }\right) $ has the property
that each $g_{q}$ is in some $H_{1}^{\beta }$. The construction of the $%
H_{i\alpha }$ from part (2) has the property that for each $\alpha $, there
exists a $\beta $ such that $H_{i\alpha }\subset H_{i}^{\beta }$ for $i=1,2$.
\end{enumerate}
\end{proof}

\begin{remark}
\ If we do not assume in part (2) that $\mathcal{H}$ is reconstructible,
then the result can be applied to the minimal reconstructible subsystem $%
H_{1c}\oplus H_{2c}$ of $\mathcal{H}$. The decoupled components $H_{id}$ can
be decomposed into $\Omega _{i}$-invariant subspaces arbitrarily, and these
are automatically $\Omega $-invariant because they are contained in $\mathrm{%
Ker}\,\Gamma ^{\dagger }$ or $\mathrm{Ker}\,\Gamma $.
\end{remark}

The first half of Theorem \ref{thmDecomposition2} generalizes to operators
with arbitrary spectrum.  We remind the reader of the standard definition of an abstract  resolution of the identitiy:

\begin{definition}[resolution of identity]
\label{resolution} Given a set $X$ and a $\sigma $-algebra $\mathcal{B}$ of
subsets of $X$ containing the empty set and $X$, we say that $\pi :\mathcal{B%
}\rightarrow \mathcal{L}(\mathcal{H})$ is a \emph{resolution of the identity}
$\mathbb{I}_{\mathcal{H}}$ if

\begin{enumerate}
\item $\pi(A)$ is an orthogonal projection for all $A \in \mathcal{B}$,

\item $\pi(\emptyset)=0$, $\pi(X)=\mathbb{I}_{\mathcal{H}}$,

\item $\pi(X\setminus A) = \mathbb{I}_{\mathcal{H}} - \pi(A)$ for all $A\in%
\mathcal{B}$,

\item $\pi(A_1\cap A_2) = \pi(A_1)\pi(A_2)$ for all $A_1,\,A_2\in\mathcal{B}$%
,

\item $\pi(A_1\cup A_2) = \pi(A_1) + \pi(A_2)$ for all $A_1,\,A_2\in\mathcal{%
B}$ with $A_1\cap A_2=\emptyset$,

\item $\pi(\bigcap\limits_{i=1}^\infty A_i) = \mathrm{s}\lim%
\limits_{i=1}^\infty \pi(A_i)$ for all sequences $\{A_i\}_{i=1}^\infty$ from 
$\mathcal{B}$ with $A_{i+1}\subset A_i$. 
\end{enumerate}

$\pi$ is said to commute with an operator $T$ if, for all $A\in\mathcal{B}$, 
$\pi(A)T = T\pi(A)$.
\end{definition}

These properties are not independent. For example, property ({\itshape v})
is implied by the first four, but we include it because of its conceptual
relevance.

The statement below is partial generalization of Theorem \ref%
{thmDecomposition} for $\Gamma $ with general spectrum.

\begin{theorem}
\label{thmDecompositionAgain} Let $\pi $ be a resolution of the identity on $%
\mathcal{H}$ that commutes with $\Omega $ and $P_{1}$. Then there exists a
resolution $E_{\pi }$ of the identity on $H_{0}$ that commutes with $\Gamma
_{0}$, such that, for all $A\in \mathcal{B}$, 
\begin{equation}
U_{i}E_{\pi }(A)=\pi (A)U_{i},\quad i=1,2,
\end{equation}%
from which it follows that 
\begin{equation}
\mathrm{Ran}\,(U_{i}E_{\pi }(A))=\mathrm{Ran}\,(\pi (A))\cap H_{i\Gamma
},\quad i=1,2.
\end{equation}
\end{theorem}


The proof of Theorem \ref{thmDecompositionAgain} is based on the following lemma.

\begin{lemma}
\label{lemmaDecomposition} Let $G_1\oplus G_2$ be s-invariant, with $%
G_1\subseteq H_1$ and $G_2\subseteq H_2$. Then there exists a $\Gamma_0$%
-invariant subspace $G_0$ of $H_{0\Gamma}$ such that 
\begin{equation}
U_i\pi_{G_0} = \pi_{G_i}U_i, \qquad i=1,2.
\end{equation}
In fact, 
\begin{equation}
G_0 = U_1^{-1} (\mathrm{Ran}\,\Gamma\cap G_1) = U_2^{-1}(\mathrm{Ran}%
\,\Gamma^\dagger\cap G_2).
\end{equation}
\end{lemma}

\begin{proof}
From Theorem \ref{thmReconstructibility2}, part ({\itshape i}), we see that 
$\mathcal{G}:=G_{1}\oplus G_{2}$ is invariant under $\mathring{\Omega}$ and
therefore also under the self-adjoint operator $\mathring{\Gamma}$; in other
words, $\mathring{\Gamma}$ commutes with the orthogonal projection $\pi _{%
\mathcal{G}}$ onto $\mathcal{G}$ in $\mathcal{H}$. Therefore, 
\begin{equation}
(\mathrm{Ran}\,\mathring{\Gamma})\cap \mathcal{G}\,=\,\mathring{\Gamma}(%
\mathcal{G})\,=\,\pi _{\mathcal{G}}(\mathrm{Ran}\,\mathring{\Gamma}),
\label{hi}
\end{equation}%
which is seen from the decomposition\ $\mathrm{Ran}\,\mathring{\Gamma}=%
\mathring{\Gamma}(\mathcal{G})\oplus \mathring{\Gamma}(\mathcal{H}\ominus 
\mathcal{G})$. From $\pi _{\mathcal{G}}=\pi _{G_{1}}\oplus \pi _{G_{2}}$ and
the definition of $\mathring{\Gamma}$, we find that \eqref{hi} admits the
decomposition 
\begin{equation}
(\mathrm{Ran}\,\Gamma )\cap G_{1}\oplus (\mathrm{Ran}\,\Gamma ^{\dagger
})\cap G_{2}\,=\,\Gamma (G_{2})\oplus \Gamma ^{\dagger }(G_{1})\,=\,\pi
_{G_{1}}(\mathrm{Ran}\,\Gamma )\oplus \pi _{G_{2}}(\mathrm{Ran}\,\Gamma
^{\dagger }).  \label{ho}
\end{equation}%
In addition, 
$\mathring{\Gamma}^{2}(\mathcal{G})=\mathring{\Gamma}(\mathcal{G})$, from
which we obtain 
\begin{equation}
\Gamma (\Gamma ^{\dagger }(G_{1}))=\Gamma (G_{2})\quad \text{and}\quad
\Gamma ^{\dagger }(\Gamma (G_{2}))=\Gamma ^{\dagger }(G_{1}).
\end{equation}%
$\mathring{\Gamma}(\mathcal{G})$ is invariant under $\mathring{\Gamma}_{R}$,
and therefore also under the unitary self-adjoint involution $\mathring{U}$
on $\mathrm{Ran}\,\mathring{\Gamma}=\mathrm{Ran}\,\Gamma \oplus \mathrm{Ran}%
\,\Gamma ^{\dagger }$ 
\begin{equation}
\mathring{U}\,:=\,\left( \mathring{\Gamma}_{R}^{2}\right) ^{-1/2}\mathring{%
\Gamma}_{R}\,,
\end{equation}%
in which we take the square root 
\begin{equation}
\left( \mathring{\Gamma}_{R}^{2}\right) ^{1/2}\,=\,\left[ 
\begin{array}{cc}
(\Gamma _{R}\Gamma _{R}^{\dagger })^{1/2} & 0 \\ 
0 & (\Gamma _{R}^{\dagger }\Gamma _{R})^{1/2}%
\end{array}%
\right] .
\end{equation}%
Using $\mathring{\Gamma}(\mathcal{G})=\Gamma (G_{2})\oplus \Gamma ^{\dagger
}(G_{1})$ and that 
\begin{equation}
\mathring{U}\,=\,\left[ 
\begin{array}{cc}
0 & U^{-1} \\ 
U & 0%
\end{array}%
\right] ,
\end{equation}%
we obtain $U\Gamma (G_{2})\subseteq \Gamma ^{\dagger }(G_{1})$ and $%
U^{-1}\Gamma ^{\dagger }(G_{1})\subseteq \Gamma (G_{2})$, so that $U\Gamma
(G_{2})=\Gamma ^{\dagger }(G_{1})$, and since $U=U_{2}U_{1}^{-1}$, we may
define 
\begin{equation}
G_{0}\,:=\,U_{1}^{-1}\Gamma (G_{2})\,=\,U_{2}^{-1}\Gamma ^{\dagger }(G_{1}).
\end{equation}%
$G_{0}$ is $\Gamma _{0}$-invariant because 
\begin{equation}
\Gamma _{0}(G_{0})=U_{1}^{-1}\Gamma U_{2}G_{0}=U_{1}^{-1}\Gamma \Gamma
^{\dagger }(G_{1})=U_{1}^{-1}\Gamma (G_{2})=G_{0}.
\end{equation}%
Finally, since $U_{1}$ takes $H_{0\Gamma }$ isomorphically to $H_{1\Gamma }$%
, we have 
\begin{equation}
U_{1}\pi _{G_{0}}\,=\,\pi _{\Gamma (G_{2})}U_{1},
\end{equation}%
in which the domain of $\pi _{\Gamma (G_{2})}$ is $H_{1\Gamma }$, and from %
\eqref{ho}, we see that $\pi _{G_{1}}$ coincides with $\pi _{\Gamma (G_{2})}$
on $H_{1\Gamma }$ so that 
\begin{equation}
U_{1}\pi _{G_{0}}\,=\,\pi _{G_{1}}U_{1}.
\end{equation}%
$U_{2}\pi _{G_{0}}=\pi _{G_{2}}U_{2}$ is obtained analogously.
\end{proof}

\vspace{1ex}

\begin{proof}[Proof of Theorem \protect\ref{thmDecompositionAgain}]
We define a map $E_{\pi }:\mathcal{B}\rightarrow \mathcal{L}(H_{0})$ and
show it is a resolution of the identity with the desired property. Let $A\in 
\mathcal{B}$ be given, and set $G_{1}=\mathrm{Ran}\,P_{1}\pi (A)$ and $G_{2}=%
\mathrm{Ran}\,P_{2}\pi (A)$. We put 
\begin{equation}
E_{\pi }(A)=\pi _{G_{0}},
\end{equation}%
where $G_{0}$ is provided by Lemma \ref{lemmaDecomposition}; the property
desired in the Theorem is thus provided by the lemma. Properties ({\itshape i%
}) and ({\itshape ii}) of a resolution of the identity are trivially
verified for $E_{\pi }$. To see property ({\itshape iii}), let $\tilde{G}%
_{0}=E_{\pi }(X\!\setminus \!A)$ and $\tilde{G}_{i}=H_{i}\ominus G_{i}=%
\mathrm{Ran}\,P_{i}\pi (X\!\setminus \!A)$, and use 
\begin{equation}
\pi _{G_{0}}=U_{i}^{-1}\pi _{G_{i}}U_{i}\quad \text{and}\quad \pi _{\tilde{G}%
_{0}}=U_{i}^{-1}\pi _{\tilde{G}_{i}}U_{i},
\end{equation}%
to calculate 
\begin{equation}
E_{\pi }(A)+E_{\pi }(X\setminus A)=\pi _{G_{0}}+\pi _{\tilde{G}%
_{0}}=U_{i}^{-1}(\pi _{G_{i}}+\pi _{\tilde{G}_{i}})U_{i}=U_{i}^{-1}\mathbb{I}%
_{H_{i}}U_{i}=\mathbb{I}_{H_{0}}.
\end{equation}%
To prove property 4 of Definion \ref{resolution}, let $A,\,B\in \mathcal{B}$%
, and compute 
\begin{multline}
E_{\pi }(A\cap B)=U_{1}^{-1}\pi (A\cap B)U_{1}=U_{1}^{-1}\pi (A)\pi (B)U_{1}
\\
=U_{1}^{-1}\pi (A)U_{1}U_{1}^{-1}\pi (B)U_{1}=E_{\pi }(A)E_{\pi }(B).
\end{multline}%
For property 5 we compute
\begin{gather}
\mathrm{s}\!\!\lim\limits_{n\rightarrow \infty }E_{\pi }(A_{n})=\mathrm{s}%
\!\!\lim\limits_{n\rightarrow \infty }U_{1}^{-1}\pi (A_{i})U_{1}=U_{1}^{-1}%
\mathrm{s}\!\!\lim\limits_{n\rightarrow \infty }(\pi (A_{n})U_{1}) \\
=U_{i}^{-1}\pi (\bigcap\limits_{n=1}^{\infty }A_{n})U_{1}=E_{\pi
}(\bigcap\limits_{n=1}^{\infty }A_{n}).  \notag
\end{gather}
\end{proof}

\vspace{2ex}
{\bfseries Acknowlegement.} \ 
The effort of A. Figotin was supported by the US Air Force Office of
Scientific Research under the Grant FA9550-04-1-0359.  The effort of S. Shipman was supported by the Louisiana Board of Regents under the Grant LEQSF(2003-06)-RD-A-14 and by the National Science Foundation under the Grant DMS-0505833.


\end{document}